\documentclass[11pt]{article}
\usepackage{amsmath,amssymb,amsfonts}
\usepackage[english]{babel}
\usepackage{float}
\makeatletter
\@addtoreset{equation}{section}
\makeatother

\def\nn{\nonumber} \def\bd{\begin{document}} \def\ed{\end{document}}
\def\ds{\documentstyle}
\let\bm=\bibitem
\newcommand{\be}{\begin{equation}}
\newcommand{\ee}{\end{equation}}
\newcommand{\bea}{\setlength\arraycolsep{2pt} \begin{eqnarray}}
\newcommand{\eea}{\end{eqnarray}}
\newcommand{\hoch}[1]{$\, ^{#1}$}
\def\p{\partial}
\title{\large {\bf Conserved quantities for asymptotically AdS spacetimes in
 quadratic curvature gravity in terms of a rank-4 tensor}}
\date{}

\author{Jun-Jin Peng$^{1,2}$\footnote{corresponding author: pengjjph@163.com},
\quad Yao Wang$^{1}$\footnote{wangytt@163.com},
\quad Wei-Jie Guo$^{1}$\footnote{wjieguo2021@163.com}\\ \\
\small \sl $^1$School of Physics and Electronic Science,
\small \sl Guizhou Normal University,\\
\small Guiyang, Guizhou 550001, People's Republic of China; \\
\small \sl  $^{2}$Guizhou Provincial Key Laboratory of Radio Astronomy and
Data Processing, \\
\small \sl Guizhou Normal University, \\
\small Guiyang, Guizhou 550001, People's Republic of China
}

\begin{document}

\maketitle
\vspace{-5pt}

\begin{center}
\textbf{Abstract}
\end{center}

We investigate the conserved quantities associated to Killing
isometries for asymptotically AdS spacetimes within the framework
of quadratic-curvature gravity. By constructing a rank-4 tensor
possessing the same index symmetries as the ones of the Riemann
tensor, we propose a 2-form potential resembling the Noether
one for quadratic-curvature gravity. Such a potential is
compared with the results via other methods existing in the
literature to establish the equivalence. Then this potential
is adopted to define conserved quantities of asymptotically
AdS spacetimes. As applications, we explicitly compute
the mass of static spherically-symmetric spacetimes, as well as
the mass and the angular momentum for rotating spacetimes,
such as the four(higher)-dimensional Kerr-AdS black holes
and black strings embedded in quadratic-curvature gravities.
Particularly, we emphasize the conserved charges of
Einstein-Gauss-Bonnet, Weyl and critical gravities,
together with the ones for the asymptotically AdS solutions
satisfying vacuum Einstein field equations.

\noindent \textbf{Keywords}: Quadratic gravity,
conserved quantity, asymptotically AdS black hole,
modified theories of gravity.

%\noindent \textbf{PACS}: 04.20.-q, 04.50.Kd, 04.70.Bw

%%%%%%%%%%%%%%%%%%%%%%%%%%%%%%%%%%%%%%%%%%%%%%%%%%%%%%%%%%%%%%%%%%%%%%%%%
\voffset=-.90pt
\vspace{10pt}
\newpage
%%%%%%%%%%%%%%%%%%%%%%%%%%%%%%%%%%
\section{Introduction}\label{one}
%%%%%%%%%%%%%%%%%%%%%%%%%%%%%%%%%%

The theories of quadratic(-curvature) gravity (it is
sometimes referred to as higher derivative gravity or
$R^2$-gravity), such as Einstein-Gauss-Bonnet gravity
\cite{RievLL,FCCM}, Weyl gravity and critical gravity
\cite{LPcritG,KKSCriG,EdNaCG}, can be regarded
as the higher-order derivative generalizations of the
well-known Einstein gravity theory. They are generally
described by the Einstein-Hilbert Lagrangian with or
without cosmological constant plus at least one of
the square of the Riemann tensor, the square of the
Ricci tensor and the quadratic Ricci curvature scalar,
or by the Lagrangian merely made up of no less than one
of the three aforementioned quadratic curvature
terms. From a mathematical perspective, such a Lagrangian
can be viewed as a functional
for the tensors of the Riemann and the metric. Since it was
discovered in \cite{Stel77} that the inclusion of quadratic
curvature terms renders the gravity theory perturbatively
renormalizable in the quantization process, quadratic-curvature
gravity has been treated as a viable candidate for a theory
of quantum gravity, and it has attracted significant
attention \cite{AKKLR,AAGS,Sal18,DonMen,BCMV,BOS92rb}.

Conventionally, to understand fully the physical and
geometric properties for a given gravity theory, a
prominent task is to search for solutions
of this theory. For quadratic-curvature gravities, in contrast
with Einstein gravity, the involvement of the quadratic curvature
terms renders it more difficult to handle the field equations.
However, if some symmetries are allowed to enter into the
metrics so that they are static and diagonal in form, it
becomes much more practicable to analytically solve the field
equations. As a consequence, a lot of static solutions with
various asymptotic structures have been found in the past
few decades. Among them, here we mention the ones presented
by the works
\cite{StatRWeyl,WeySsMK,BouDes,GBWhe,GBWil,GBBHcai,GBCNO,NojOd02,Nels10,
LPPS15,KKLR15,BueCa,KKZhi,NojOd17,BonS19,SPPP,PSPP20,PPPS21,PACMo,PPOr23,
HLZ23,Nguy23,Ngu232,ANg23,BAdSEGB,GSMah,HyNam,BLifBJV,ABJR22,BriRa,
PPS17,GGST11,GST12}.
On the other hand, although it is of great difficulty to
construct exact rotating solutions in quadratic-curvature
gravities without systematical methods generating solutions,
one feasible way to achieve this is to embed the rotating
solutions obeying the vacuum Einstein field equations into
such theories. For example, the four-dimensional stationary
and axially-symmetric Kerr-AdS black hole solution
\cite{4DKeAdSsolu} is likewise the one for
Einstein-Gauss-Bonnet, Weyl and critical gravities.
Besides, it will be demonstrated below that some
quadratic-curvature gravities
are able to embrace the higher-dimensional generalizations of
the four-dimensional Kerr-AdS solution \cite{GLuPP1,GLuPP2}.
With those solutions in hand, as usual, a necessary procedure is
to give their conserved charges for the sake of understanding
thermodynamic properties, which are of considerable interest
at the present stage. As a matter of fact, there exist
a number of approaches for conserved charges in the literature,
such as
the covariant phase space method \cite{LeW,IyW,WZA} and its
development \cite{BarnB}, the Abbott-Deser-Tekin (ADT) formalism
\cite{AbbottD,AbbottD2,DeserT,DeserT2,KimKY},
the Ashtekar-Magnon-Das (AMD) method \cite{AMDmass,AMDmass2}
and the field-theoretic approach \cite{PeEGB,PetNDT,PetPit}.
In the last few years, in the spirit of these methods or
other ones, many efforts have been given from diverse perspectives
to explore the
conserved quantities of various spacetimes in
quadratic-curvature gravities, particularly the asymptotically
AdS ones \cite{PinGR,PinGR1,PLnovCC,AmGor,SeST,DeSar,Bay4QG,MOPCouT,
WanPen,GMOR,GMORB,Kas08,Ort21,GOKAMD,PaAMD,ASSTquad,JJPEPJC,FanL15,AMOcm}.

Within this paper, in spite of the fact that a lot of methods
have been devoted to the definition for the conserved quantities
of quadratic-curvature gravities, we attempt to propose a simple
and convenient formulation of conserved quantities for
asymptotically AdS spacetimes within the framework of these
theories and then make use of the formulation to explicitly
illustrate how the quadratic curvature terms correct the mass
and the angular momentum of black holes in the context of
Einstein gravity.
For this purpose, we shall follow the work \cite{PLnovCC} to
construct a rank-4 tensor that exhibits the same index
symmetries with those of the Riemann tensor. The linear
combination of such a tensor with another rank-4 one that is
defined by the derivative of the Lagrangian with respect
to the Riemann tensor
further gives rise to a 2-form potential associated to an
arbitrary Killing vector field, which takes a similar structure
as the Noether one. Of particular interest will be the
applications of this potential in Einstein-Gauss-Bonnet,
Weyl and critical gravities. By virtue of the comparison of the
potential with the results via other methods, it will be
demonstrated that its integral on a codimension-2 surface
can bring about an appropriate formula for conserved charges
of asymptotically AdS spacetimes according to Stokes' theorem.
Furthermore, by utilizing this formula, we shall
calculate the mass for static and spherically symmetric
spacetimes, as well as the mass and the angular momentum
of four(higher)-dimensional Kerr-AdS black holes corrected by
terms in quadratic curvatures.

The rest of the present paper is organised as follows. In section \ref{two},
starting with the general form for the Lagrangian of quadratic-curvature
gravities, we shall introduce a rank-4 tensor to construct the potentials
associated to conserved quantities of these theories. Such potentials will
be compared with the ones via other methods in the literature, and their
applications in some typical quadratic-curvature gravities will be
strengthened. In section \ref{three}, we will apply the formula for
conserved charges to compute the mass of static spherically-symmetric
spacetimes with the asymptotically AdS structure, including the ones
in general relativity and Einstein-Gauss-Bonnet gravity in arbitrary
dimensions, as well as the ones in four-dimensional Weyl and critical
gravities. In section \ref{four}, we shall compute the mass and
angular momenta of four(higher)-dimensional rotating Kerr-AdS
black holes and black strings embedded into the theories of
quadratic-curvature gravity. The last section is devoted to
our conclusions.

%%%%%%%%%%%%%%%%%%%%%%%%%%%%%%%%%%%%%%%%%%%%%%%%%%%%%%%%%%%%%%%%%%%%%%%%
\section{The general formalism}\label{two}
%%%%%%%%%%%%%%%%%%%%%%%%%%%%%%%%%%%%%%%%%%%%%%%%%%%%%%%%%%%%%%%%%%%%%%%%

In this section, we shall investigate potentials defined in terms
of a rank-4 tensor with the same index symmetries as those of
the Riemann tensor within the framework of the theories of
quadratic-curvature gravity. It will be demonstrated that
such potentials are equivalent to the ones via other methods,
such as the (off-shell) ADT formalism, the covariant phase
space approach, the generalized Komar integral and the
field-theoretic method. As a consequence, it is allowed to apply them to
define conserved quantities of asymptotically AdS spacetiems in
these gravity theories. In particular, we are going to analyze
the applications of the potentials in general relativity,
Einstein-Gauss-Bonnet gravity, Weyl gravity and critical gravity.

In the present work, the integer $D$ stands for the dimensions of
spacetimes. We adopt the notations in \cite{WaldGR} to define
the Riemann curvature tensor $R_{\mu\nu\rho\sigma}$ through
$\big(\nabla_\mu\nabla_\nu-\nabla_\nu\nabla_\mu\big)V_\rho
=R_{\mu\nu\rho\sigma}V^\sigma$(here $V_\rho$ denotes an arbitrary
vector field), while $R_{\mu\nu}=g^{\rho\sigma}R_{\rho\mu\sigma\nu}$
and $R=g^{\rho\sigma}R_{\rho\sigma}$ represent the Ricci tensor
and its scalar curvature respectively. For generality, we take
into consideration of the usual Einstein-Hilbert Lagrangian
in the presence of a negative cosmological constant $\Lambda$
plus the linear combination for all the quadratic
curvature terms $R^2$, $R^{\alpha\beta}R_{\alpha\beta}$
and $R^{\mu\nu\rho\sigma}R_{\mu\nu\rho\sigma}$, written
as the following form
\be
\sqrt{-g}L=\sqrt{-g}\big(R-2\Lambda+c_1R^2
+c_2R^{\alpha\beta}R_{\alpha\beta}
+c_3R^{\mu\nu\rho\sigma}R_{\mu\nu\rho\sigma}\big)
\, .\label{QuaCLag}
\ee
Here $(c_1,c_2,c_3)$ represent coupling constants. With different choices
of these constants, Eq. (\ref{QuaCLag}) admits the Lagrangians for
some typical quadratic-curvature gravities. For example, it includes
Weyl gravity as a special case, which is regarded as a compelling
alternative to Einstein gravity. Specifically, the Weyl tensor in
$D$-dimensional spacetime is given by \cite{WaldGR}
\begin{equation}
C^{\mu\nu}_{\rho\sigma}=
\frac{R}{(D-1)(D-2)}
\delta^{\mu\nu}_{\rho\sigma}
-\frac{4}{D-2}R^{[\mu}_{[\rho}\delta^{\nu]}_{\sigma]}
+R^{\mu\nu}_{\rho\sigma}
\, .\label{WeylTen}
\end{equation}
Here and in what follows, we follow the convention whereby
the square brackets enclosing indices
denote anti-symmetrization of them, such that
$R^{[\mu}_{[\rho}\delta^{\nu]}_{\sigma]}=
(R^{[\mu}_{\rho}\delta^{\nu]}_{\sigma}
-R^{[\mu}_{\sigma}\delta^{\nu]}_{\rho})/2$, and the generalized
Kronecker delta symbol
$\delta^{\mu_1\cdot\cdot\cdot\mu_m}_{\nu_1\cdot\cdot\cdot\nu_m}
=m!\delta^{[\mu_1}_{[\nu_1}\cdot\cdot\cdot\delta^{\mu_m]}
_{\nu_m]}$. An arbitrary rank-4 tensor
$X^{\mu\nu}_{\rho\sigma}
=g^{\mu\alpha}g^{\nu\beta}X_{\alpha\beta\rho\sigma}$.
It is easy to check that the Weyl tensor
$C^{\mu\nu}_{\rho\sigma}$
is traceless, namely, $C^{\mu\rho}_{\nu\rho}=C^{\rho\mu}_{\rho\nu}=0$.
The contraction between two Weyl tensors is read off as
\bea
C^{\mu\nu}_{\rho\sigma}C_{\mu\nu}^{\rho\sigma}
&=&\frac{1}{4}\delta^{\gamma\lambda\mu\nu}_{\alpha\beta\rho\sigma}
C^{\alpha\beta}_{\gamma\lambda}C_{\mu\nu}^{\rho\sigma}
=C^{\mu\nu}_{\rho\sigma}R_{\mu\nu}^{\rho\sigma}
\nonumber  \\
&=&\frac{2}{(D-1)(D-2)}R^2
-\frac{4}{D-2}R^{\rho}_{\sigma}R_{\rho}^{\sigma}
+R^{\mu\nu}_{\rho\sigma}R_{\mu\nu}^{\rho\sigma}
\, .\label{CCcontr}
\eea
For the purpose to obtaining the first equality in
Eq. (\ref{CCcontr}), we have made use of the expansion
for the generalized Kronecker delta symbol
$\delta^{\gamma\lambda\mu\nu}_{\alpha\beta\rho\sigma}$, that is,
\bea
\delta^{\gamma\lambda\mu\nu}_{\alpha\beta\rho\sigma}&=&
\delta^{\gamma\lambda}_{\alpha\beta}\delta^{\mu\nu}_{\rho\sigma}
-\delta^{\gamma\mu}_{\alpha\beta}\delta^{\lambda\nu}_{\rho\sigma}
-\delta^{\gamma\nu}_{\alpha\beta}\delta^{\mu\lambda}_{\rho\sigma}
\nonumber  \\
&&-\delta^{\mu\lambda}_{\alpha\beta}\delta^{\gamma\nu}_{\rho\sigma}
-\delta^{\nu\lambda}_{\alpha\beta}\delta^{\mu\gamma}_{\rho\sigma}
+\delta^{\mu\nu}_{\alpha\beta}\delta^{\gamma\lambda}_{\rho\sigma}
\, .  \label{Rank4Kron}
\eea
In the absence of the $R-2\Lambda$ term, together with all the three
coupling constants satisfying $c_2=-2(D-1)c_1$ and $c_3=(D-1)(D-2)c_1/2$,
the Lagrangian (\ref{QuaCLag}) becomes the one for $D$-dimensional
Weyl gravity, namely,
\be
\sqrt{-g}L_W=\frac{c_1(D-1)(D-2)}{2}
\sqrt{-g}C^{\mu\nu\rho\sigma}C_{\mu\nu\rho\sigma}
\, ,\label{WeylLag}
\ee
where Eq. (\ref{CCcontr}) has been used. Besides, when $c_2=-4c_1$
and $c_3=c_1$, Eq. (\ref{QuaCLag}) transforms into the Lagrangian for
Einstein-Gauss-Bonnet gravity \cite{RievLL,FCCM}, taking the form
\begin{equation}
\sqrt{-g}L_{EGB}=\sqrt{-g}\big(R-2\Lambda+c_1L_{GB}\big)
\, ,\label{GBLag}
\end{equation}
in which the Gauss-Bonnet invariant $L_{GB}$ is read off as
\begin{equation}
L_{GB}=R^2-4R^{\alpha\beta}R_{\alpha\beta}
+R^{\mu\nu\rho\sigma}R_{\mu\nu\rho\sigma}
=\frac{1}{4}\delta^{\gamma\lambda\mu\nu}_{\alpha\beta\rho\sigma}
R^{\alpha\beta}_{\gamma\lambda}R_{\mu\nu}^{\rho\sigma}
\, . \label{GBterm}
\end{equation}
In order to arrive at the last equality in Eq. (\ref{GBterm}),
we have utilized Eq. (\ref{Rank4Kron}). As a consequence of
the Gauss-Bonnet-Chern theorem (see Refs.
\cite{GilPa,DerRot,LJAlt} for this theorem in the context
of pseudo-Riemann manifolds and its implications in
gravity theories), the integration of
the Gauss-Bonnet invariant on $D=4$ compact manifold
gives rise to a constant with a value relying on the
four-dimensional Euler characteristic of the manifold.
When $D=4$, the Gauss-Bonnet term is often referred to as
a topological invariant, the addition of which to the
Lagrangian makes no contribution to
the modification of the bulk dynamics. However, this term
is of great importance in the renormalization of the
Einstein gravity theory. Besides, it can be
adopted to define the conserved charges in asymptotically
(locally) AdS spaces within the context of the four-dimensional
Einstein gravity or quadratic gravity \cite{GMORB,ACOTZ}.
What is more,
due to the vanishing of the generalized Kronecker delta symbol
$\delta^{\gamma\lambda\mu\nu}_{\alpha\beta\rho\sigma}$ in
three dimensions, Eq. (\ref{GBterm}) leads to the identity
$R^2-4R^{\alpha\beta}R_{\alpha\beta}
+R^{\mu\nu\rho\sigma}R_{\mu\nu\rho\sigma}=0$ or
$C^{\mu\nu}_{\rho\sigma}C_{\mu\nu}^{\rho\sigma}=0$
in three dimensions, which can be reproduced by means
of using the equation $C^{\mu\nu}_{\rho\sigma}(D=3)=0$
derived from the expansion of the identity
$\delta^{\gamma\lambda\mu\nu}_{\alpha\beta\rho\sigma}
R^{\alpha\beta}_{\gamma\lambda}=0$ $(D=3)$ in terms of
Eq. (\ref{Rank4Kron}). The substitution of the
three-dimensional vacuum Einstein field equations
$R_{\mu\nu}=2\Lambda g_{\mu\nu}$ into
$C^{\mu\nu}_{\rho\sigma}(D=3)=0$ yields
$R^{\mu\nu}_{\rho\sigma}=-\Lambda
\delta^{\mu\nu}_{\rho\sigma}$.
This implies that all the vacuum solutions in
three-dimensional Einstein gravity with the
cosmological constant are locally equivalent
to the ones with maximal symmetries,
whose Riemann curvature tensor obeys the relation
$R^{\mu\nu}_{\rho\sigma}\propto\delta^{\mu\nu}_{\rho\sigma}$.

To connect more straightforwardly the Lagrangian (\ref{QuaCLag})
with some typical quadratic gravities, when the dimension $D\geq4$,
with Eqs. (\ref{CCcontr}) and (\ref{GBterm}), the Lagrangian
(\ref{QuaCLag}) can be reexpressed as the form consisting of
the usual Einstein-Hilbert part, the square of the Weyl tensor,
the Gauss-Bonnet invariant and the quadratic Ricci curvature scalar
term, namely,
\bea
L&=&R-2\Lambda+\frac{(D-2)(c_2+4c_3)}{4(D-3)}
C^{\mu\nu}_{\rho\sigma}C_{\mu\nu}^{\rho\sigma} \nonumber \\
&&-\frac{(D-2)c_2+4c_3}{4(D-3)}L_{GB}
+\left(c_1+\frac{Dc_2+4c_3}{4D-4}\right)R^2
\, . \label{QuaCLag2}
\eea
One can take advantage of Eq. (\ref{QuaCLag2}) in the description for
some specifical quadratic-curvature gravities. For example, within the
case where $D=4$, $c_2=-3c_1$ and $c_3=0$, Eq. (\ref{QuaCLag2})
gives rise to the Lagrangian $\mathcal{L}^{(4D)}_{CG}$
for four-dimensional critical gravity \cite{LPcritG}, having the form
\begin{equation}
\mathcal{L}^{(4D)}_{CG}=\sqrt{-g}\Big[R-2\Lambda
+\frac{3}{2}c_1\big(L_{GB}
-C^{\mu\nu}_{\rho\sigma}C_{\mu\nu}^{\rho\sigma}\big)\Big]
\, . \label{4DCriG}
\end{equation}
Here the constant parameter $c_1$ is imposed to take the specifical value
$c_1=-1/(2\Lambda)$. It should be pointed out that the expression
(\ref{4DCriG}) differs from the original one for the Lagrangian given
by \cite{LPcritG}. Such an expression renders it convenient to
reveal the relationships among critical gravity, Weyl gravity and
Einstein-Gauss-Bonnet gravity, as well as to achieve the
higher-dimensional generalization according to Eq. (\ref{LCriGinD})
in Appendix \ref{appendB}. Apart from the theory of critical gravity
characterized by the Lagrangian (\ref{4DCriG}),
neglecting the $R-2\Lambda$ part and letting $c_1=3\alpha$,
$c_2=-12\alpha$ and $c_3=6\alpha$ in Eq. (\ref{QuaCLag2}),
where $\alpha$ represents an arbitrary constant parameter, one acquires
another type of four-dimensional critical gravity proposed in terms of the
four-dimensional scale invariant gravity in \cite{EdNaCG}, whose Lagrangian
is the linear combination of the one for Weyl gravity with a quadratic
Ricci curvature scalar $R^2$ term, namely,
\be
\tilde{\mathcal{L}}^{(4D)}_{CG}=\alpha \sqrt{-g}\Big(R^2
+6 C^{\mu\nu}_{\rho\sigma}C_{\mu\nu}^{\rho\sigma}\Big)
=-3\alpha \sqrt{-g}\Big(R^2
-4R^{\rho\sigma}R_{\rho\sigma}\Big)
+6\alpha \sqrt{-g}L_{GB}
\, . \label{4DCriG2}
\ee
It will be demonstrated in Appendix \ref{appendA} that
the Lagrangian (\ref{4DCriG2}) allows for the existence
of asymptotically AdS solutions, and its
higher-dimensional generalization will be given
by Eq. (\ref{DDCriG2}) in Appendix \ref{appendB}.
In addition, when $c_1=-D^2/[8\Lambda(D-2)^2]$, $c_2=-4(D-1)c_1/D$
and $c_3=0$, the Lagrangian (\ref{QuaCLag}) or (\ref{QuaCLag2}) becomes
the one in Eq. (\ref{LCriGinD}), which can be thought of as the
higher-dimensional generalization of the Lagrangian (\ref{4DCriG})
for four-dimensional critical gravity \cite{KKSCriG}.

Next, we take into account the variation of the Lagrangian
(\ref{QuaCLag}) with respect to the metric tensor $g_{\mu\nu}$.
We write down
\be
\delta\big(\sqrt{-g}L\big)=\sqrt{-g}E_{\mu\nu} \delta g^{\mu\nu}
+\sqrt{-g}\nabla_\mu \Theta^\mu
\, . \label{Lagvari}
\ee
In the above equation, the expression for field equations
$E_{\mu\nu}$ is presented by \cite{Pady}
\bea
E_{\mu\nu} &=&\left(\frac{\partial L}{\partial g^{\mu\nu}}\right)
_{R_{\bullet\bullet\bullet\bullet}}
-\frac{1}{2}Lg_{\mu\nu}
-R_{(\mu}^{~~~\lambda\rho\sigma}P_{\nu)\lambda\rho\sigma}
 -2\nabla^\rho\nabla^\sigma P_{\rho(\mu\nu)\sigma}\nonumber  \\
&=&R_{(\mu}^{~~~\lambda\rho\sigma}P_{\nu)\lambda\rho\sigma}
 -2\nabla^\rho\nabla^\sigma P_{\rho(\mu\nu)\sigma}
 -\frac{1}{2}Lg_{\mu\nu} \nn \\
 &=&R_{\mu}^{~~\lambda\rho\sigma}P_{\nu\lambda\rho\sigma}
 -2\nabla^\rho\nabla^\sigma P_{\rho\mu\nu\sigma}
 -\frac{1}{2}Lg_{\mu\nu}
\, . \label{MotionEq}
\eea
Here $R_{\bullet\bullet\bullet\bullet}$ stands for the covariant
rank-4 Riemann curvature tensor. The expression
$\big(\partial L/\partial g^{\mu\nu}\big)_{R_{\bullet\bullet\bullet\bullet}}
=2R_{(\mu}^{~~~\lambda\rho\sigma}P_{\nu)\lambda\rho\sigma}$
\cite{Pady,KluLi,EdRR} and Eq. (\ref{Rm2delP}) have been used to gain
the second and the last  equalities, respectively. The surface term
$\Theta^\mu$ in Eq. (\ref{Lagvari}) is written as
\begin{equation}
\Theta^\mu
=2P^{\mu\nu\rho\sigma}\nabla_\sigma\delta g_{\rho\nu}
-2\delta g_{\nu\rho} \nabla_\sigma P^{\mu\nu\rho\sigma}
\, .\label{Boudterm}
\end{equation}
In Eqs. (\ref{MotionEq}) and (\ref{Boudterm}), the tensor
$P^{\mu\nu\rho\sigma}$ is defined in terms of the derivative
with respect to the Riemann tensor, that is,
$P^{\mu\nu}_{\rho\sigma}
=\big(\partial L/\partial R^{\rho\sigma}_{\mu\nu}
\big)_{g_{\alpha\beta},g^{\gamma\lambda}}$, which is
referred to as entropy tensor in \cite{Pady}, attributed
to the fact that its integral on the Killing horizon
can give rise to the entropy of some higher-order derivative
modified gravity theories \cite{IyW}. By virtue of
Eqs. (\ref{ParRRR}) and (\ref{ParRRR1}) in Appendix
\ref{appendA}, the tensor
$P^{\mu\nu}_{\rho\sigma}$ takes the following form
\bea
P^{\mu\nu}_{\rho\sigma}&=&\left(
\frac{\partial L}{\partial R^{\rho\sigma}_{\mu\nu}}
\right)_{g_{\alpha\beta},g^{\gamma\lambda}}
\nonumber  \\
&=&\frac{1}{2}\delta^{\mu\nu}_{\rho\sigma}
+c_1R\delta^{\mu\nu}_{\rho\sigma}
+2c_2R^{[\mu}_{[\rho}\delta^{\nu]}_{\sigma]}
+2c_3R^{\mu\nu}_{\rho\sigma}
\, . \label{PRtensor}
\eea
Particularly, within the context of Weyl and Einstein-Gauss-Bonnet
gravities, the tensor $P^{\mu\nu}_{\rho\sigma}$ is represented by
$P^{\mu\nu}_{W\rho\sigma}$ and $P^{\mu\nu}_{EGB\rho\sigma}$,
respectively, which are expressed as
\bea
P^{\mu\nu}_{W\rho\sigma}
&=&\frac{\partial L_W}{\partial R^{\rho\sigma}_{\mu\nu}}
=c_1(D-1)(D-2)C^{\mu\nu}_{\rho\sigma} \, , \nonumber  \\
P^{\mu\nu}_{EGB\rho\sigma}
&=&\frac{\partial L_{EGB}}{\partial R^{\rho\sigma}_{\mu\nu}}
=\frac{1}{2}\delta^{\mu\nu}_{\rho\sigma}
+\frac{c_1}{2}\delta^{\gamma\lambda\mu\nu}_{\alpha\beta\rho\sigma}
R^{\alpha\beta}_{\gamma\lambda}
\, . \label{PWGBten}
\eea
Substituting Eqs. (\ref{PRtensor}) and (\ref{TwodivP}) into
Eq. (\ref{MotionEq}), we further write down the expression
for the field equations $E_{\mu\nu}$, being of the form
\bea
E_{\mu\nu}&=&R_{\mu\nu}+2c_1RR_{\mu\nu}
+2(c_2+2c_3)R_{\mu\rho\nu\sigma}R^{\rho\sigma}
\nonumber  \\
&&-4c_3R_{\mu\lambda}R^\lambda_\nu
+2c_3R_{\mu}^{~~\lambda\rho\sigma}R_{\nu\lambda\rho\sigma}
-\frac{1}{2}Lg_{\mu\nu}
 \nonumber  \\
&&+\frac{1}{2}(4c_1+c_2)g_{\mu\nu}\Box R
+(c_2+4c_3)\Box R_{\mu\nu}
\nonumber  \\
&&-(2c_1+c_2+2c_3)\nabla_\mu\nabla_\nu R
\, .  \label{MotionEq2}
\eea
Here $E_{\mu\nu}$ can be also found in the works
\cite{BCMV,FanL15,BHLif,GulTek} and Eq. (\ref{DivEmn})
demonstrates that $E_{\mu\nu}$ is divergence-free.
It should be pointed out that $E_{\mu\nu}$ might shed
light on the understanding for the field equations of
higher-order gravities as it has been demonstrated
in \cite{BCMV} that the linearized expressions
for the equations of motion in any higher-order
gravity characterized by the Lagrangian made up of
the Riemann tensor can be always mapped to those
in the quadratic-curvature gravity theory.

With the expression $E_{\mu\nu}$ for the field equations in hand,
let us explore its AdS space solutions before proceeding any
further. To do so, it is assumed that the Lagrangian (\ref{QuaCLag})
admits $D$-dimensional maximally symmetric AdS spacetimes with
the line element
\be
d\bar{s}^2 = -\big(1-\hat{\Lambda}r^2\big)dt^2
+\frac{dr^2}{1-\hat{\Lambda}r^2}+r^2d\Omega_{D-2}^2
\, . \label{LEofAdS}
\ee
Here and in what follows, quantities with an overline refer to the AdS
background metric (\ref{LEofAdS}). In the above expression, $t$ and $r$
represent the time and radial coordinate respectively,
$d\Omega_{D-2}^2$ denotes the line element for the $(D-2)$-dimensional
unit sphere, and the constant parameter $\hat{\Lambda}$ could be
thought of as an effective cosmological constant with the general form
$\hat{\Lambda}=\hat{\Lambda}(\Lambda,c_1,c_2,c_3)$. According to
Eq. (\ref{LEofAdS}), the Riemann curvature
tensor $\bar{R}^{\mu\nu}_{\rho\sigma}$, the Ricci curvature tensor
$\bar{R}^{\mu}_{\rho}$ and scalar $\bar{R}$ are respectively given by
\begin{equation}
\bar{R}^{\mu\nu}_{\rho\sigma}=\hat{\Lambda}\delta^{\mu\nu}_{\rho\sigma}
\, , \quad
\bar{R}^{\mu}_{\rho}=(D-1)\hat{\Lambda}\delta^{\mu}_{\rho}\, , \quad
\bar{R}=D(D-1)\hat{\Lambda}
\, .\label{AdScurv}
\end{equation}
The substitution of Eq. (\ref{AdScurv}) into Eq. (\ref{MotionEq2}) results
in the expression $\bar{E}^\mu_\nu$ for the field equations of the AdS
spacetimes, being of the form
\bea
\bar{E}^\mu_\nu&=&\bar{R}_{\rho\sigma}^{\mu\lambda}
\bar{P}_{\nu\lambda}^{\rho\sigma}
 -\frac{1}{4}\delta^\mu_\nu
 \bar{R}_{\alpha\beta}^{\rho\sigma}
 \bar{P}^{\alpha\beta}_{\rho\sigma}
 -\frac{1}{4}\bar{R}\delta^\mu_\nu+\Lambda \delta^\mu_\nu
 \nonumber  \\
&=&-\frac{1}{4}\big[(D-1)(2Dk-8k+D)\hat{\Lambda}
-4\Lambda\big] \delta^\mu_\nu
\, . \label{EforAdS}
\eea
To simplify the calculations, in the first equality of
Eq. (\ref{EforAdS}), we have made use of the identity
given by Eq. (\ref{Laltexpre}). The constant $k$ in
Eq. (\ref{EforAdS}) is defined through the value
of the tensor $P^{\mu\nu}_{\rho\sigma}$ on the
$D$-dimensional AdS spacetimes, that is,
\be
\bar{P}^{\mu\nu}_{\rho\sigma}
=P^{\mu\nu}_{\rho\sigma}(g_{\alpha\beta}\rightarrow
\bar{g}_{\alpha\beta})
=k\delta^{\mu\nu}_{\rho\sigma}
\, , \label{PRbar}
\ee
and it is read off as
\be
k=\frac{1}{2}+[(D-1)(Dc_1+c_2)+2c_3]\hat{\Lambda}
\, . \label{kvalue}
\ee
For instance, $k=1/2$ for Einstein gravity, $k=0$ for both
Weyl gravity in any dimension and four-dimensional
critical gravity described by the Lagrangian (\ref{4DCriG}),
and $k_{EGB}=k(c_2=-4c_1,c_3=c_1)$
for Einstein-Gauss-Bonnet gravity. For the purpose to
guaranteeing that the Lagrangian (\ref{QuaCLag}) allows for
the AdS space solution (\ref{LEofAdS}), it is required that
$\bar{E}^\mu_\nu=0$. In this regard, Eq. (\ref{EforAdS}) shows
that the constant parameter $\hat{\Lambda}$ has to be constrained
by the following condition
\cite{BCMV,PACMo,MOPCouT,ASSTquad,FanL15}
\be
\frac{D-4}{D-2}[(D-1)(Dc_1+c_2)+2c_3]\hat{\Lambda}^2
+\hat{\Lambda}-\hat{\Lambda}_{gr}=0
\, , \label{HatLamcons}
\ee
in which the constant $\hat{\Lambda}_{gr}$ is defined by
\be
\hat{\Lambda}_{gr} = \frac{2\Lambda}{(D-1)(D-2)}
\, . \label{Lamgr}
\ee
Furthermore, by the aid of Eq. (\ref{kvalue}), the constraint
(\ref{HatLamcons}) is reexpressed as
\begin{equation}
(D-4)k=(D-2)\frac{\hat{\Lambda}_{gr}}{\hat{\Lambda}}
-\frac{D}{2}
\, . \label{HatLamcons2}
\end{equation}
According to Eq. (\ref{HatLamcons}) or Eq. (\ref{HatLamcons2}),
$\hat{\Lambda}=\hat{\Lambda}_{gr}$ for Einstein gravity. In the case of
Weyl gravity, the constraint $(D-1)(Dc_1+c_2)=-2c_3$ leads to
$\hat{\Lambda}=\hat{\Lambda}_{gr}$. As a result, the $D$-dimensional
AdS spacetimes in general relativity are also the solutions for
Weyl gravity. In the case for Einstein-Gauss-Bonnet gravity, the
constraint (\ref{HatLamcons}) is rewritten as
$\tilde{c}_1\hat{\Lambda}^2+\hat{\Lambda}-\hat{\Lambda}_{gr} =0$.
Here and in what follows, $\tilde{c}_1=(D-3)(D-4)c_1$. Its solutions
are $\hat{\Lambda}
=\big[-1\pm(1+4\tilde{c}_1\hat{\Lambda}_{gr})^{1/2}\big]
/(2\tilde{c}_1)$. What is more, in $D=4$ dimensions,
$\hat{\Lambda} =\hat{\Lambda}_{gr} = \Lambda/3$ holds for all
four-dimensional quadratic-curvature gravities. This means that
the four-dimensional AdS space is an exact solution of all
these gravities.

In the remainder of the present section, we are going to follow the work
\cite{PLnovCC} to investigate the definition of the conserved quantities
associated to Killing isometries for asymptotically AdS spacetimes
in the framework of quadratic-curvature gravity theories depicted
generally by the Lagrangian (\ref{QuaCLag}). For this purpose,
a crucial procedure is to construct a rank-4 tensor
$P^{\mu\nu}_{(ref)\rho\sigma}$ inheriting the index symmetries
of the Riemann curvature tensor, which is of the form
\begin{equation}
P^{\mu\nu}_{(ref)\rho\sigma} = \frac{1}{4(D-3)\hat{\Lambda}}
\delta^{\gamma\lambda\mu\nu}_{\alpha\beta\rho\sigma}
R^{\alpha\beta}_{\gamma\lambda}
-\frac{D-4}{2}\delta^{\mu\nu}_{\rho\sigma}
\, .  \label{Preften}
\end{equation}
It can be proven that $P^{[\mu\nu\rho]\sigma}_{(ref)}=0$ and
$P^{\mu\nu\rho\sigma}_{(ref)}$ is conserved identically,
namely, $\nabla_\mu P^{\mu\nu\rho\sigma}_{(ref)}=0$.
Specially, for the case of the AdS spacetime (\ref{LEofAdS}),
one observes that
$\bar{P}^{\mu\nu}_{(ref)\rho\sigma}
=P^{\mu\nu}_{(ref)\rho\sigma}\big|_{g=\bar{g}}
=\delta^{\mu\nu}_{\rho\sigma}$. Moreover, through the linear
combination of both the tensors $P^{\mu\nu}_{\rho\sigma}$
and $P^{\mu\nu}_{(ref)\rho\sigma}$, another rank-4 tensor
$\mathcal{P}^{\mu\nu}_{\rho\sigma}$ is defined as
\be
\mathcal{P}^{\mu\nu}_{\rho\sigma}
=P^{\mu\nu}_{\rho\sigma}-kP^{\mu\nu}_{(ref)\rho\sigma}
\, .  \label{CalPdef}
\ee
Apparently, the tensor $\mathcal{P}^{\mu\nu}_{\rho\sigma}$
inherits all the index symmetries of the Riemann curvature
tensor and it disappears on the AdS spacetime, namely,
$\mathcal{P}^{\mu\nu}_{\rho\sigma}\big|_{g=\bar{g}}=0$.
It is further assumed that the asymptotically AdS spacetimes
admits the symmetry generated by a Killing vector field
$\xi^\mu$, which can be associated to a conserved current
$J^\mu$ in accordance with Noether theorem. According to
Poincare lemma, $J^\mu$ corresponding to the Killing
vector $\xi^\mu$ is read off as
$J^\mu=\nabla_\nu K^{\mu\nu}$,
where the 2-form potential $K^{\mu\nu}$ is
proposed as \cite{PLnovCC}
\be
K^{\mu\nu} = \mathcal{P}^{\mu\nu}_{\rho\sigma}\nabla^{\rho}\xi^{\sigma}
-2\xi^\sigma \nabla^{\rho}\mathcal{P}^{\mu\nu}_{\rho\sigma}
\, ,  \label{Kmunu0}
\ee
in terms of the rank-4 tensor $\mathcal{P}^{\mu\nu}_{\rho\sigma}$.
Here we point out that $K^{\mu\nu}$ is our suggested potential that
is appropriate for the definition of conserved charges of
asymptotically AdS spacetimes within the framwork of
quadratic-curvature gravities. By the aid of Eq. (\ref{TwodivP}),
the equations of motion $E^\mu_\nu=0$ and the identity
$\nabla_\mu \nabla_\nu \xi_\rho=R_{\rho\nu\mu\sigma}\xi^\sigma$
for the Killing vector $\xi^\mu$, the substitution of
Eq. (\ref{Kmunu0}) into the expression for the conserved current
$J^\mu$ yields
\bea
J^\mu&=& \xi^\nu\big(P^{\mu\lambda}_{\rho\sigma}R^{\rho\sigma}_{\nu\lambda}
-2\nabla^\rho\nabla^\sigma P^\mu_{~\rho\sigma\nu}\big)
+kP^{\mu\nu}_{(ref)\rho\sigma}R^{\rho\sigma}_{\nu\kappa}\xi^\kappa\nn \\
&=&\frac{1}{2}L\xi^\mu+\frac{k}{4(D-3)\hat{\Lambda}}
\delta^{\gamma\lambda\mu\nu}_{\alpha\beta\rho\sigma}
R^{\alpha\beta}_{\gamma\lambda}
R^{\rho\sigma}_{\nu\kappa}\xi^\kappa
+(D-4)kR^{\mu}_{\nu}\xi^\nu
\, . \label{ConCurJ}
\eea
Here the second equality is achieved under the on-shell condition
for the metric tensor. With the help of
$\nabla_\mu(L\xi^\mu)=\xi^\mu\nabla_\mu L=0$,
$2\nabla_\mu(R^{\mu}_{\nu}\xi^\nu)=2R_{\mu\nu}\nabla^\mu\xi^\nu
+\xi^\mu\nabla_\mu R=0$ and
$\sqrt{-g}\nabla_\mu
\big(\delta^{\gamma\lambda\mu\nu}_{\alpha\beta\rho\sigma}
R^{\alpha\beta}_{\gamma\lambda}
R^{\rho\sigma}_{\nu\kappa}\xi^\kappa\big)
=\partial_\mu\partial_\nu\big(\sqrt{-g}
\delta^{\gamma\lambda\mu\nu}_{\alpha\beta\rho\sigma}
R^{\alpha\beta}_{\gamma\lambda}
\nabla^{\rho}\xi^{\sigma}\big)=0$, one can verify
that $J^\mu$ is conserved, namely, $\nabla_\mu J^\mu=0$.
At the same time, one observes that $J^\mu$ vanishes on the
AdS spacetimes, namely, $J^\mu|_{g=\bar{g}}=\bar{J}^\mu=0$,
or equivalently, the exterior derivative for the Hodge dual of
the two-form potential $K^{\mu\nu}$ fulfills
$({\rm d}\star K)|_{g=\bar{g}}=0$.
On the other hand, as usual, by means of the variation for the
Lagrangian (\ref{QuaCLag}) with respect to the metric tensor,
together with the Lie derivative with regard to the diffeomorphism
symmetry generated by the Killing vector field $\xi^\mu$, one obtains
the 2-form Noether potential $K^{\mu\nu}_R$, given by
\begin{equation}
K^{\mu\nu}_R=P^{\mu\nu}_{\rho\sigma}\nabla^{\rho}\xi^{\sigma}
-2\xi^\sigma \nabla^{\rho}P^{\mu\nu}_{\rho\sigma}
\, .\label{KRmunu}
\end{equation}
Obviously, $K^{\mu\nu}$ resembles the Noether potential
$K^{\mu\nu}_R$ due to the fact that the replacement of
the tensor $P^{\mu\nu\rho\sigma}$ with
$\mathcal{P}^{\mu\nu\rho\sigma}$ in Eq. (\ref{KRmunu})
makes $K^{\mu\nu}_R$ coincide with $K^{\mu\nu}$ and both
the tensors $P^{\mu\nu\rho\sigma}$ and
$\mathcal{P}^{\mu\nu\rho\sigma}$ have the same index symmetries.
In terms of $K^{\mu\nu}_R$, the potential $K^{\mu\nu}$ can
be further expressed as an alternative form
\be
K^{\mu\nu} = K^{\mu\nu}_R
-kP^{\mu\nu}_{(ref)\rho\sigma}\nabla^{\rho}\xi^{\sigma}
\, .  \label{KmunuR}
\ee
In this respect, the potential $K^{\mu\nu}$ can be
decomposed into two components. The first one is
the usual Noether potential $K^{\mu\nu}_R$, while
the second one $kP^{\mu\nu}_{(ref)
\rho\sigma}\nabla^{\rho}\xi^{\sigma}$ is
responsible for curing divergences appearing in
the calculations of $K^{\mu\nu}_R$. The existence
of the factor $\hat{\Lambda}^{-1}$ in the rank-4
tensor $P^{\mu\nu}_{(ref)\rho\sigma}$ may render
the second component divergent when
$\hat{\Lambda}\rightarrow 0$. However, since
the other quantities $K^{\mu\nu}_R$,
$R^{\mu\nu}_{\rho\sigma}$ and
$\nabla^{\rho}\xi^{\sigma}$ involved in
the potential $K^{\mu\nu}$ generally depend
on the parameter $\hat{\Lambda}$ for asymptotically
AdS spacetimes, their total contribution can
guarantee the convergence of $K^{\mu\nu}$ under the
limit $\hat{\Lambda}\rightarrow 0$. This situation
happens also to the potentials via the
topological regulation method \cite{GMOR,GMORB},
as well as to the potential for four-dimensional
Einstein gravity proposed in \cite{ACOTZ}, since
such potentials explicitly contain the inverse of
the cosmological constant. Therefore, the above
suggests that it is feasible to take the limit
$\hat{\Lambda}\rightarrow 0$ on the potential
$K^{\mu\nu}$ so as to extend $K^{\mu\nu}$ to compute 
the conserved charges of asymptotically flat 
counterparts for asymptotically AdS spacetimes.

Remarkably, the structure
of $K^{\mu\nu}$ is similar as the Komar-type potentials
for the theories of higher-order derivative gravity proposed
in \cite{Kas08,Ort21}. In fact, according to these works,
those Komar-type potentials can be generally
expressed as the form $K^{\mu\nu}_{gK}=K^{\mu\nu}_R-B^{\mu\nu}$
with the anti-symmetric tensor $B^{\mu\nu}$ defined through
$\nabla_\nu B^{\mu\nu}=1/2L\xi^\mu$. Here the 2-form $B^{\mu\nu}$
is determined up to a divergence-free 2-form, and its
local existence is always guaranteed attributed to the
fact that the divergence of $L\xi^\mu$ vanishes identically for
the diffeomorphism invariant Lagrangian $L$ and the Killing
vector field $\xi^\mu$. The divergence
$\nabla_\nu K^{\mu\nu}_R=1/2L\xi^\mu$ under the on-shell
condition, cancelling out the divergence of the 2-form
$B^{\mu\nu}$ in the conserved current
$J^{\mu}_{gK}=\nabla_\nu K^{\mu\nu}_{gK}$. As a consequence,
one obtains $J^{\mu}_{gK}=0$. In this regard, both the currents
$J^{\mu}$ and $J^{\mu}_{gK}$ coincides with each other on the
AdS spacetimes, rendering it of possibility for the integrals
of $K^{\mu\nu}$ and $K^{\mu\nu}_{gK}$ on the codimension-2
surfaces at infinity to yield the same asymptotic charges.
Due to the above, the
$kP^{\mu\nu}_{(ref)\rho\sigma}\nabla^{\rho}\xi^{\sigma}$
ingredient in $K^{\mu\nu}$ can be interpreted as a
prospective substitution for the 2-form $B^{\mu\nu}$
at infinity, and the potential $K^{\mu\nu}$ could be
regarded as a Komar-like potential for the theories of
quadratic-curvature gravity.

Before any further process, here we give three significant
examples on the applications of the potential $K^{\mu\nu}$.
As a direct application to the AdS spacetime (\ref{LEofAdS}),
$K^{\mu\nu}$ turns into the one
$\bar{K}^{\mu\nu}=K^{\mu\nu}|_{g=\bar{g}}=0$, implying that
$K^{\mu\nu}$ vanishes identically on the AdS spacetime.
In addition, when the potential (\ref{Kmunu0}) is applied
to the theory of Weyl gravity described by the
Lagrangian (\ref{WeylLag}), one substitutes
the tensor $P^{\mu\nu}_{W\rho\sigma}$ in
Eq. (\ref{PWGBten}) into the Noether potential
$K^{\mu\nu}_R$ to acquire the
potential $K^{\mu\nu}_{Weyl}$, given by
\be
K^{\mu\nu}_{Weyl}=c_1(D-1)(D-2)
\big(C^{\mu\nu}_{\rho\sigma}\nabla^{\rho}\xi^{\sigma}
-2\xi^\sigma \nabla^{\rho}C^{\mu\nu}_{\rho\sigma}\big)
\, ,\label{KmunuW}
\ee
which is just the Noether potential for Weyl gravity, arising from
that $k=0$. It can be tested that the perturbation of the potential
$K^{\mu\nu}_{Weyl}$ on four-dimensional AdS spaces is
equivalent to the one given by Eq. (23) in \cite{JJPEPJC},
which was acquired via the (off-shell)
ADT formalism \cite{AbbottD,AbbottD2,DeserT,DeserT2,KimKY}.
Moreover, for the $D$-dimensional Einstein-Gauss-Bonnet
gravity with the Lagrangian (\ref{GBLag}), its potential
$K^{\mu\nu}_{EGB}$, derived from Eq. (\ref{Kmunu0}) without the
on-shell condition for the metric, takes the following form
\be
K^{\mu\nu}_{EGB} = \frac{1+2\tilde{c}_1
\hat{\Lambda}}{2}\left[(D-2)\nabla^{\mu}\xi^{\nu}
-\frac{1}{4(D-3)\hat{\Lambda}}
\delta^{\gamma\lambda\mu\nu}_{\alpha\beta\rho\sigma}
R^{\alpha\beta}_{\gamma\lambda}
\nabla^{\rho}\xi^{\sigma}\right]
\, ,  \label{KmnEGB}
\ee
which is proportional to the potential for Einstein
gravity given in \cite{PinGR,PinGR1,PLnovCC} by the
factor $\big(1+2\tilde{c}_1\hat{\Lambda}\big)$ regardless
of the on-shell condition. As a matter of fact, since it
will be demonstrated below that the linear
perturbation of $K^{\mu\nu}$ with respect to
the decomposition to the metric
$g_{\mu\nu}=\delta g_{\mu\nu}+\bar{g}_{\mu\nu}$,
where $\bar{g}_{\mu\nu}$ is the metric of the AdS
space (\ref{LEofAdS}), is consistent with the
potential defined by the ADT formalism, the same
perturbation for $K^{\mu\nu}_{EGB}$, being a
special case of $K^{\mu\nu}$, is equivalent to the
superpotential (3.11) on the AdS space proposed
in \cite{PeEGB}. The latter was obtained via the
field-theoretic approach \cite{PeEGB,PetNDT,PetPit} and
was verified to coincide with the ADT potential
in \cite{PeEGB}. Apart from this, the linear
perturbation of $K^{\mu\nu}_{EGB}$ on AdS spaces
is equivalent to the ADT potential given by Eq. (41)
in \cite{JJPEPJC}. However, $K^{\mu\nu}_{EGB}$ here is
much simpler than the superpotentials in
\cite{PeEGB,JJPEPJC}. In particular,
under a critical condition of Eq. (\ref{KmnEGB}),
where $1+2\tilde{c}_1\hat{\Lambda}=0$ or both the
constant parameters $c_1$ and $\Lambda$ are related
to each other through
\begin{equation}
c_1=-\frac{(D-1)(D-2)}{8(D-3)(D-4)\Lambda}
\, ,  \label{Lc1EGB}
\end{equation}
the potentials for $D(D\geq5)$-dimensional Einstein-Gauss-Bonnet
gravities vanish identically, leading to zero conserved charges.

For convenience to calculations, with the help of the expression
(\ref{Rank4Kron})
for the expansion of the generalized Kronecker-delta symbol
$\delta^{\gamma\lambda\mu\nu}_{\alpha\beta\rho\sigma}$, as well as
Eq. (\ref{divPR}) for divergence of the rank-4 tensor
$P^{\mu\nu}_{\rho\sigma}$, the potential $K^{\mu\nu}$ given by
Eq. (\ref{Kmunu0}) is rewritten as
\bea
K^{\mu\nu} &=& \big[1+(D-4)k+(2c_1-\hat{k})R\big]\nabla^{[\mu}\xi^{\nu]}
\nonumber  \\
&&-2\big(c_2+2\hat{k}\big)R^{[\mu}_\rho\nabla^{\nu]}\xi^{\rho}
+\big(2c_3-\hat{k}\big)R^{\mu\nu}_{\rho\sigma}\nabla^\rho\xi^\sigma
\nonumber  \\
&&+(4c_1+c_2)\xi^{[\mu}\nabla^{\nu]}R
-2(c_2+4c_3)\xi^\rho\nabla^{[\mu}R^{\nu]}_\rho
\, . \label{Kmunu}
\eea
Here the constant parameter $\hat{k}$ is presented by
\begin{equation}
\hat{k}=\frac{k}{(D-3)\hat{\Lambda}}
\, .
\end{equation}
It is worth to mentioning that the potential $K^{\mu\nu}$ in
Eq. (\ref{Kmunu}) can go further under the on-shell condition
that the metric tensor is the solution of the field equations
$E_{\mu\nu}=0$. In terms of the following identities associated
with the Killing vector $\xi^{\nu}$
\begin{align}
\Box\nabla^{\mu}\xi^{\nu} &=
-R^{\mu\nu}_{\rho\sigma}\nabla^\rho\xi^\sigma
-\xi^\rho\nabla^{[\mu}R^{\nu]}_\rho
\, , \nonumber  \\
\nabla^{\mu}\Box\xi^{\nu} &=
-\xi^\rho\nabla^{\mu}R^{\nu}_\rho
-R^{\nu}_\rho\nabla^{\mu}\xi^{\rho}
\, , \label{Killpro}
\end{align}
the potential $K^{\mu\nu}$ in Eq. (\ref{Kmunu}) can be reformulated
in the language of differential forms into an alternative form
\bea
K&=&\frac{1}{2}\big[1+(D-4)k+(6c_1+c_2-\hat{k})R\big]
{\rm d}\xi
\nonumber \\
&&-\Big(c_2+2\hat{k}\Big){\rm d}\Box\xi
+2\big(c_2+2c_3+\hat{k}\big)\Box {\rm d}\xi
\nonumber \\
&&+\left(2c_2+5c_3+\frac{3}{2}\hat{k}\right)
R_{\mu\nu}^{\rho\sigma}\nabla_\rho\xi_\sigma
{\rm d}x^\mu\wedge {\rm d}x^\nu
\nonumber \\
&&-\frac{1}{2}(4c_1+c_2){\rm d}(R\xi)
\, . \label{Kmndf}
\eea
Eq. (\ref{Kmndf}) demonstrates that the potential $K^{\mu\nu}$
could be reproduced through the action of the differential
operators on the Killing vector field. This can be seen more
clearly due to the one-form current $J=-\star {\rm d}\star K$
\cite{PZL}. In comparison, the differential form (\ref{Kmndf})
for the potential $K^{\mu\nu}$ can be regarded as a special
case of the general potential proposed in \cite{PenZ}.

We move on to present the concrete expressions for the potential
$K^{\mu\nu}$ given by Eq. (\ref{Kmunu}) in some special cases
of the Lagrangian (\ref{QuaCLag}). When it becomes the 
Einstein-Hilbert one $L_{EH}=R-(D-1)(D-2)\hat{\Lambda}$ with
$\hat{\Lambda}=\hat{\Lambda}_{gr}$ under $c_1=c_2=c_3=0$, 
by the aid of the equations
of motion $R_{\mu\nu}=(D-1)\hat{\Lambda}g_{\mu\nu}$, the potential
$K^{\mu\nu}$ is simplified as \footnote{The potential
$K_{gr}^{\mu\nu}$ can be interpreted as the generalization
of the usual Komar potential $\nabla^{[\mu}\xi^{\nu]}$ in
asymptotically AdS spacetimes. In contrast with the
latter, the former incorporates an additional
second-order derivative term
$(2\hat{\Lambda})^{-1}
R^{\mu\nu}_{\rho\sigma}\nabla^\rho\xi^\sigma$. Such a
term was also introduced in the potential given by
Eq. (8) in \cite{ACOTZ}, which was proposed to define
the conserved charges of four-dimensional Einstein gravity
with cosmological constant. The $(2\hat{\Lambda})^{-1}
R^{\mu\nu}_{\rho\sigma}\nabla^\rho\xi^\sigma$ term
plays the main role in eliminating the divergent terms
appearing within
$\nabla^{[\mu}\xi^{\nu]}$ to render a finite result.
Apart from this, it corrects the normalization factors
in the explicit calculations for the conserved quantities
of spacetimes. In particular, $K_{gr}^{\mu\nu}$ vanishes
on the AdS spaces in Eq. (\ref{LEofAdS}). Additionally,
for ultrastatic spacetimes with the
metric ansatz $ds^2=-dt^2+g_{ij} dx^idx^j$, where the 
Riemannian metric $g_{ij}$ on the $(D-1)$-dimensional space is 
independent of the time coordinate $t$ \cite{UltraSS}, 
due to the fact that $R^t_{~\rho\mu\nu}=0$ 
and the conserved quantities are defined via the integral 
of $K_{gr}^{\mu\nu}$ over the surface at 
$t=\text{Const}$, the $(2\hat{\Lambda})^{-1}
R^{\mu\nu}_{\rho\sigma}\nabla^\rho\xi^\sigma$ term 
actually makes no contribution to the conserved 
quantities. Thus, $K_{gr}^{\mu\nu}$ is equivalent 
to the Komar potential, rendering it feasible to 
use $K_{gr}^{\mu\nu}$ to yield the Komar
charges for asymptotically-flat ultrastatic spacetimes.
Since the timelike Killing vector $\xi^\mu=-\delta^\mu_t$
of such spacetimes obeys $\nabla^\mu \xi^\nu=0$, further
yielding $K_{gr}^{\mu\nu}=0$, the energy
of ultrastatic spacetimes vanishes.}
\be
K_{gr}^{\mu\nu}=\frac{1}{D-3}
\Big(\nabla^{[\mu}\xi^{\nu]}-\frac{1}{2\hat{\Lambda}}
R^{\mu\nu}_{\rho\sigma}\nabla^\rho\xi^\sigma\Big)
\, , \label{Kgrgen}
\ee
which is consistent with the usual Komar potential
modified by an additional second-order derivative
term proportional to the Riemann
curvature tensor, given by Eq. (2.15) in \cite{PZL}.
When $c_3=0$, the potential
$K^{\mu\nu}$ in $D=4$ dimensions coincides with the one
given by Eq. (16) in \cite{WanPen} or Eq. (26) in
\cite{GMOR}, and $K^{\mu\nu}$ in $D=2(n+2)$ dimensions
is equivalent to the potential given by Eq. (23) in
\cite{WanPen} or Eq. (14) in \cite{GMORB}, acquired
via the topological regularization method.
When the Lagrangian (\ref{QuaCLag}) admits the solutions
obeying $R_{\mu\nu}=(D-1)\hat{\Lambda}g_{\mu\nu}$, for such solutions,
the potential $K^{\mu\nu}$ takes the form
\begin{equation}
K_{Ric}^{\mu\nu}=2\Big[k-2(D-3)c_3\hat{\Lambda}\Big]K_{gr}^{\mu\nu}
\, . \label{KRic}
\end{equation}
In particular, when $D=4$, one obtains
$K_{Ric}^{\mu\nu}\big|_{D=4}
=2\big[1+6(4c_1+c_2)\hat{\Lambda}\big]
K_{gr}^{\mu\nu}\big|_{D=4}$,
implying that the potential is irrelevant to the coupling
constant $c_3$ or the scalar term
$R^{\mu\nu}_{\rho\sigma}R_{\mu\nu}^{\rho\sigma}$.
Obviously, the potential $K_{Ric}^{\mu\nu}$ is proportional to
$K_{gr}^{\mu\nu}$, supporting the finding that the ADT
potentials of quadratic curvature gravities are proportional
to the one for Einstein gravity in \cite{PinGR1,ASSTquad}.
At the linearized level, the potential $K_{Ric}^{\mu\nu}$
completely agrees with the one given by
Eq. (6) in \cite{SeST}, and the conserved charge
defined in terms of $K_{Ric}^{\mu\nu}$ is consistent with the one
given by Eq. (2.3) in \cite{AmGor}. What is more, in the case of
the four-dimensional critical gravity depicted by the Lagrangian
(\ref{4DCriG}), the constraints $c_2=-3c_1$ and $c_3=0$
enable us to express $K^{\mu\nu}$ as a simpler form
\bea
K^{\mu\nu}_{4DCG} &=& \nabla^{[\mu}\xi^{\nu]}
-\frac{1}{2\Lambda}\Big(2R\nabla^{[\mu}\xi^{\nu]}
+6R^{[\mu}_\rho\nabla^{\nu]}\xi^{\rho}
+\xi^{[\mu}\nabla^{\nu]}R
+6\xi^\rho\nabla^{[\mu}R^{\nu]}_\rho\Big)
\nonumber  \\
&=& -3\nabla^{[\mu}\xi^{\nu]}
-\frac{3}{\Lambda}\Big(R^{[\mu}_\rho\nabla^{\nu]}\xi^{\rho}
+\xi^\rho\nabla^{[\mu}R^{\nu]}_\rho\Big)
\, , \label{Kmn4DCG}
\eea
or equivalently formulated in the language of exterior algebra
as follows
\bea
K_{4DCG}&=&-\frac{3}{2}{\rm d}\xi
-\frac{3}{2\Lambda}{\rm d}\Box\xi
+\frac{3}{\Lambda}\Box {\rm d}\xi
+\frac{3}{\Lambda}R_{\mu\nu}^{\rho\sigma}\nabla_\rho\xi_\sigma
{\rm d}x^\mu\wedge {\rm d}x^\nu \nonumber \\
&=&\frac{1}{2}{\rm d}\xi
-\frac{3}{2\Lambda}{\rm d}\Box\xi
+\frac{3}{\Lambda}\Box {\rm d}\xi
-4K_{gr}
\, . \label{Kmn4DCG2}
\eea
In Eq. (\ref{Kmn4DCG}), we have used the result $R=4\Lambda$
derived from Eq. (\ref{4DLonshel}) to arrive at the second
equality. It can be verified that the perturbation of the potential
$K^{\mu\nu}_{4DCG}$ on the AdS spacetime (\ref{LEofAdS}) vanishes.

In the above, some special cases of the potential $K^{\mu\nu}$
have been compared with the ones via other methods (the results
will be summarized in Appendix \ref{appendD}). Apart from this,
for generality, it is of great necessity to build the relation
between $K^{\mu\nu}$ and the well-known Iyer-Wald potential defined
by the covariant phase space method \cite{LeW,IyW,WZA} in the context
of the Lagrangian (\ref{QuaCLag}). By virtue of following the
covariant phase space approach, the Iyer-Wald potential associated
with the Killing vector $\xi^\mu$ is read off as \cite{BCMV}
\begin{equation}
Q^{\mu\nu}_{IW}=\delta K^{\mu\nu}_R
+\frac{1}{2}K^{\mu\nu}_Rg^{\rho\sigma}\delta g_{\rho\sigma}
-\xi^{[\mu}\Theta^{\nu]}
\, ,\label{IWpoten}
\end{equation}
which coincides with the off-shell generalized ADT potential
\cite{KimKY}. In Eq. (\ref{IWpoten}), the surface term $\Theta^{\mu}$
and  the Noether potential
$K^{\mu\nu}_R$ are given by Eqs. (\ref{Boudterm}) and (\ref{KRmunu})
respectively. As a special case of (\ref{IWpoten}), where the
background spacetime is the fixed $D$-dimensional AdS space (\ref{LEofAdS})
and the linear perturbation of the metric tensor is defined through
$\delta g_{\mu\nu}=g_{\mu\nu}-\bar{g}_{\mu\nu}$,
the potential $Q^{\mu\nu}_{IW}$ becomes
\bea
\bar{Q}^{\mu\nu}_{IW}&=&\delta K^{\mu\nu}_R
+\frac{1}{2}g^{\alpha\beta}\delta g_{\alpha\beta}
\bar{P}^{\mu\nu}_{\rho\sigma}
\bar{\nabla}^\rho\bar{\xi}^\sigma
-2\bar{\xi}^{[\mu}\bar{P}^{\nu]\lambda}_{\rho\sigma}
\bar{g}^{\rho\alpha}\bar{\nabla}^\sigma
\delta g_{\alpha\lambda}
\nonumber  \\
&=&\delta K^{\mu\nu}_R
+k\bar{g}^{\alpha\beta}\delta g_{\alpha\beta}
\bar{\nabla}^{[\mu}\bar{\xi}^{\nu]}
-2k\bar{\xi}^{[\mu}\delta^{\nu]\lambda}_{\rho\sigma}
\bar{g}^{\rho\alpha}\bar{\nabla}^\sigma
\delta g_{\alpha\lambda}
\, , \label{QbarIW}
\eea
which is consistent with the ADT potential for the Lagrangian
(\ref{QuaCLag}) \cite{DeserT,DeserT2,ASSTquad}. On the other hand,
with the help of Eq. (\ref{QbarIW}),
the linear perturbation of the potential $K^{\mu\nu}$ on the AdS
space (\ref{LEofAdS}) gives rise to the following result
\begin{equation}
\delta K^{\mu\nu}=\bar{Q}^{\mu\nu}_{IW}
-\hat{k}\bar{\nabla}_\gamma \bar{U}^{\gamma\mu\nu}
\, . \label{QdelK}
\end{equation}
In Eq. (\ref{QdelK}), the 3-form $\bar{U}^{\gamma\mu\nu}$ is presented by
\begin{equation}
\bar{U}^{\gamma\mu\nu}=\frac{1}{2}
\delta^{\lambda\gamma\mu\nu}_{\alpha\beta\rho\sigma}
\bar{g}^{\beta\eta}\big(\bar{\nabla}^\alpha\delta g_{\eta\lambda}\big)
\bar{\nabla}^\rho\bar{\xi}^\sigma
\, . \label{Ubardef}
\end{equation}
As a consequence of Eq. (\ref{QdelK}), the perturbation of the potential
$K^{\mu\nu}$ on the AdS spacetime background is equivalent to the
Iyer-Wald potential $\bar{Q}^{\mu\nu}_{IW}$ on the same background,
as well as to the ADT potential. This is attributed to the fact that the
integral of ${\rm d}\star \bar{U}$, which stands for the exterior
derivative for the Hodge dual of the 3-form $\bar{U}$, vanishes
according to Stokes' theorem. In this regard, the
$kP^{\mu\nu}_{(ref)\rho\sigma}\nabla^{\rho}\xi^{\sigma}$
component in $K^{\mu\nu}$ plays the role of compensating the
contribution from the $\xi^{[\mu}\Theta^{\nu]}$ part in Iyer-Wald
potential $Q^{\mu\nu}_{IW}$. So the potential
$K^{\mu\nu}$ could be thought of as the ``integration form" of
$Q^{\mu\nu}_{IW}$ on the fixed AdS background. Moreover, the perturbation
of the potential $K^{\mu\nu}$ on the AdS spacetime background is
equivalent to the ones presented in \cite{PinGR1,AmGor,SeST,DeSar}.
The potential $K^{\mu\nu}$ in four dimensions or in odd dimesions
is perhaps equivalent to the one proposed in \cite{Bay4QG} in the
absence of torsion, or the one obtained via counterterm method
in \cite{MOPCouT} respectively.

At the end, we are going to make use of the potential $K^{\mu\nu}$
given by Eq. (\ref{Kmunu0}) or (\ref{Kmunu}) to define the conserved
charges of quadratic-curvature gravities described by the
Lagrangian (\ref{QuaCLag}). To achieve this, as usual, it is
assumed that there exists a $(D-1)$-dimensional hypersurface
$\Sigma$ with the boundary $\partial\Sigma$. In terms of a
$(D-2)$-form $\star K$, which is the Hodge dual of the 2-form
potential $K^{\mu\nu}$, according to Stokes' theorem, a formula
for the conserved charges of such gravity theories can be put
forward as the integral of the potential $\star K$ over the
$(D-2)$-dimensional surface $\partial\Sigma$, that is,
\be
Q = \frac{1}{8\pi}\int_{\partial\Sigma}
\star K
\, . \label{Qdef}
\ee
Here let us make some remarks on the formula (\ref{Qdef}) for
conserved quantities. Firstly, according to the equivalence
relation displayed by Eq. (\ref{QdelK}), the potential
$K^{\mu\nu}$ can be used to replace the ADT and the Iyer-Wald
ones when the latter two are applied to compute the
conserved quantities of asymptotically AdS spacetimes
in general relativity and quadratic gravities.
However, in contrast with the conserved
quantities by means of the latter two, the formula (\ref{Qdef})
has the merit of avoiding the computation for the perturbation
of the potential, greatly simplifying the calculations.
Secondly, $K^{\mu\nu}$ resembles the Noether potential
$K^{\mu\nu}_R$ as well as the potential
$K^{\mu\nu}_{gK}$. Accordingly, the formula (\ref{Qdef})
takes a similar structure as the Komar-type integral
and the Wald entropy formula, providing a simple and
convenient formulation for the conserved quantities
of asymptotically AdS spacetimes in quadratic gravities.
However, unlike the original Komar integral, which
suffers the disadvantage to present an expression for
the mass differing from that for the angular momentum
by an anomalous factor in form, the formula (\ref{Qdef})
gives a unified form of the conserved charges.
Thirdly, Eq. (\ref{KmunuR}) shows that the potential
$K^{\mu\nu}$ can be decomposed into two parts. One is
the Noether potential $K^{\mu\nu}_R$, while the other is
the anti-symmetric rank-2 tensor
$-kP^{\mu\nu}_{(ref)\rho\sigma}\nabla^{\rho}\xi^{\sigma}$. However,
when the formula (\ref{Qdef}) is adopted to compute the mass
of asymptotically AdS spacetimes, it can not be split into
two parts according to the decomposition of $K^{\mu\nu}$,
attributed to the fact that each separate part of
the formula suffers from divergence at spatial infinity
although their combination is convergent.

%%%%%%%%%%%%%%%%%%%%%%%%%%%%%%%%%%%%%%%%%%%%%%%%%%%%%%%%%%%%%%%%%%%%%%%%
\section{Mass for static and spherically symmetric spacetimes}\label{three}
%%%%%%%%%%%%%%%%%%%%%%%%%%%%%%%%%%%%%%%%%%%%%%%%%%%%%%%%%%%%%%%%%%%%%%%%

In this section, as an application for the formula of conserved charges,
we shall compute the mass for general static and spherically symmetric
spacetimes endowed with an asymptotically AdS structure. Such spacetimes are
solutions obeying the field equation $E_{\mu\nu}=0$, and they cover the
ones in general relativity, four-dimensional Weyl and critical
gravities, and Einstein-Gauss-Bonnet gravity in some special cases.

We begin with the general metric ansatz for $D$-dimensional asymptotically
AdS spacetimes with spherical symmetry.
In the coordinate system $\{t,r,y^i\}$ $(i=1,2,\cdot\cdot\cdot,D-2)$,
where $y^i$s denote the coordinates parametrizing a $(D-2)$-dimensional
unit sphere, the line element of spactime can be always expressed as
the following general form
\bea
ds^2 &=& -f(r)dt^2+\frac{dr^2}{f(r)}+B^2(r)d\Omega_{D-2}^2 \, ,\nonumber  \\
d\Omega_{D-2}^2&=&h_{ij}(y)dy^idy^j
\, .\label{SSlineele}
\eea
In the above expression, $h_{ij}$ represents the
metric tensor for the codimension-2 unit sphere.
However, here we point out that all the results
related to the curvature tensors below hold for
an arbitrary codimension-2 spacial compact
manifold. In order to maintain the asymptotically
AdS structure (\ref{LEofAdS}), the functions
$f(r)$ and $B(r)$ in Eq. (\ref{SSlineele}) are
required to behave at the asymptotic infinity as
$f(\infty)\rightarrow 1-\hat{\Lambda}r^2$ and
$B(\infty)\rightarrow r$, respectively. Moreover,
under the coordinate transformation
$\varrho=\varrho(r)$ determined by $\varrho=B(r)$,
the line element (\ref{SSlineele}) is recast
into another common form:
\be
ds^2 = -F(\varrho)dt^2+\frac{d\varrho^2}{H(\varrho)}
+\varrho^2d\Omega_{D-2}^2
\, .\label{SSlinrho}
\ee
Here the functions $F(\varrho)$ and $H(\varrho)$
are given respectively by
\be
F(\varrho)=f[r(\varrho)] \, , \quad
H(\varrho)=f[r(\varrho)]\left(\frac{dB}{dr}\right)^{2}
\, . \label{FHrho}
\ee
In terms of the line element (\ref{SSlineele}) or (\ref{SSlinrho}),
there have been a lot of works devoting to seeking exact
static and spherically symmetric solutions within the context of
higher-derivative gravity theories. For example, see the works
\cite{StatRWeyl,WeySsMK,BouDes,GBWhe,GBWil,GBBHcai,GBCNO,NojOd02,Nels10,
LPPS15,KKLR15,BueCa,KKZhi,NojOd17,BonS19,SPPP,PSPP20,PPPS21,PACMo,PPOr23}
and related references therein.

As an attempt to acquire the concrete expressions for the mass of the
general static spherically-symmetric black holes in terms of the
formula (\ref{Qdef}) for the conserved charges, it is of great
necessity to compute the relevant curvature tensors for the
spacetimes described by Eq. (\ref{SSlineele}). Implementing
computations to the $(t,r,\rho,\sigma)$ components of the Riemann
curvature tensor $R^{\mu\nu}_{\rho\sigma}$ gives rise to
\begin{equation}
R^{tr}_{\rho\sigma}=-f^{\prime\prime}\delta^t_{[\rho}\delta^r_{\sigma]}
\, .\label{Riemtr}
\end{equation}
Here and in what follows, the quantity with the prime ``$\prime$"
denotes its derivative with respect to the radial coordinate $r$, such as
$f^{\prime}=df/dr$ and $f^{\prime\prime}=d^2f/dr^2$. Furthermore,
after some complicated calculations, we obtain the related components
of the Ricci tensor, which are read off as $R_{ti}=R_{ri}=0$, together
with the ones
\bea
R^t_t &=& -\frac{1}{2B}\big[Bf^{\prime\prime}
+(D-2)B^{\prime}f^{\prime}\big]\, ,\nonumber  \\
R^r_r&=&-\frac{1}{2B}\big[Bf^{\prime\prime}
+(D-2)\big(2fB^{\prime\prime}+B^{\prime}f^{\prime}\big)\big]
\, ,\nonumber  \\
R^j_i&=&\frac{1}{B^2}\big[R^j_{hi}-\delta^j_i\big(BfB^{\prime\prime}
+BB^{\prime}f^{\prime}+(D-3)fB^{\prime2}\big)\big]
\, . \label{Rictr}
\eea
In the above equation, $R^j_{hi}$ represents the Ricci curvature
tensor for the $(D-2)$-dimensional line element $d\Omega_{D-2}^2$,
whose Ricci curvature scalar is defined through $R_{h}=h^{ij}R_{hij}$.
On the basis of Eq. (\ref{Rictr}), the Ricci curvature scalar $R$,
defined as $R=R^\mu_\mu=R^t_t+R^r_r+R^i_i$, is presented by
\begin{equation}
R=\frac{1}{B^2}\big[R_h-B^2f^{\prime\prime}-(D-2)\big(2BfB^{\prime\prime}
+2BB^{\prime}f^{\prime}+(D-3)fB^{\prime2}\big)\big]
\, .\label{Ricsca}
\end{equation}
Particularly, in the case where $h_{ij}$ is the metric tensor for
the $(D-2)$-dimensional unit sphere, the Ricci curvature scalar
$R_h$ in Eq. (\ref{Ricsca}) takes the value $R_h=(D-2)(D-3)$.

With Eqs. (\ref{Riemtr}), (\ref{Rictr}) and (\ref{Ricsca}) in hand,
wo process to compute the 2-form potentials in terms of $K^{\mu\nu}$
given by Eq. (\ref{Kmunu}). As usual, the Killing vector corresponding
to the mass for the static and spherically symmetric spacetimes
characterized by the line element (\ref{SSlineele}) is  chosen as
$\xi^\mu=(-1,0,\cdot\cdot\cdot,0)$. With such a Killing vector, we calculate
the $tr$ component of the potential $K^{\mu\nu}$, yielding
\bea
K^{tr} &=&\frac{c_2+2\hat{k}}{2}\big(R^t_t+R^r_r\big)f^{\prime}
\nonumber  \\
&&-\frac{2c_3-\hat{k}}{2}f^{\prime}f^{\prime\prime}
-\frac{4c_1+c_2}{2}f\partial_r R  \nonumber  \\
&&+\frac{f^{\prime}}{2}\big[1+(D-4)k+(2c_1-\hat{k})R\big]
\nonumber  \\
&&-\frac{c_2+4c_3}{2B}\big[2Bf\partial_r R^t_t
+(D-2)ff^{\prime}B^{\prime\prime}\big]
\, . \label{Ktrcompts}
\eea
Subsequently, we take into account several special aspects of $K^{tr}$.
First, when the Lagrangian (\ref{QuaCLag}) turns into the one
for Einstein gravity, in which $c_1=c_2=c_3=0$ and $k=1/2$, the
component $K^{tr}$ coincides with the one given by a
generalized Komar formulation \cite{PZL}. It is simplified as
\be
K^{tr}_{gr}=\frac{D-2}{4}\left(f^{\prime}
-\frac{f^{\prime}}{\hat{\Lambda}B^2}
+\frac{ff^{\prime}B^{\prime2}}{\hat{\Lambda}B^2}\right)
\, . \label{Kgrtr}
\ee
Second, in the framework of Einstein-Gauss-Bonnet gravity,
where $c_2=-4c_1$ and $c_3=c_1$, the $tr$ component $K^{tr}$ is
expressed as
\bea
K^{tr}_{EGB}&=& \frac{f^{\prime}}{2}\big[1+(D-4)k_{EGB}\big]
\nonumber  \\
&&-\frac{k_{EGB}-2c_1(D-3)\hat{\Lambda}}{2(D-2)^{-1}\hat{\Lambda}B^2}
\big(1-fB^{\prime2}\big)f^{\prime}\nonumber  \\
&=&\big[1+2c_1(D-3)(D-4)\hat{\Lambda}\big]K^{tr}_{gr}
\, . \label{KtrEGB}
\eea
In the above equation, the constant $k_{EGB}$ is determined by
plugging $c_2=-4c_1$ and $c_3=c_1$ into Eq. (\ref{kvalue}),
being of the form $k_{EGB}=1/2+(D-2)(D-3)c_1\hat{\Lambda}$. Apparently,
the second equality in Eq. (\ref{KtrEGB}) verifies the observation
in Eq. (\ref{KmnEGB}) that the potential for Einstein-Gauss-Bonnet
gravity is proportional to the one for Einstein gravity. Third, in the
case for the four-dimensional Weyl gravity characterized by
the Lagrangian (\ref{WeylLag}), the component $K^{tr}$ becomes
\begin{equation}
K^{tr}_{4DWG}= \frac{c_1}{B^3}K^{tr}_{qc}
\, . \label{Ktr4DW}
\end{equation}
In Eq. (\ref{Ktr4DW}), the quantity $K^{tr}_{qc}$ represents the contribution
from all the quadratic curvature terms in the Lagrangian (\ref{WeylLag}),
being of the form
\bea
K^{tr}_{qc}&=&2Bf^{\prime}-4fB^{\prime}
-6Bff^{\prime}B^{\prime2}  \nonumber  \\
&&+4f^2B^{\prime}B^{\prime}B^{\prime}+\big(2ff^{\prime\prime\prime}
-f^{\prime}f^{\prime\prime}\big)B^3\nonumber  \\
&&+2\big(fB^{\prime}f^{\prime\prime}-2f^2B^{\prime\prime\prime}
-3ff^{\prime}B^{\prime\prime}+B^{\prime}f^{\prime2}\big)B^2
\, . \label{KtrQuaC}
\eea
Fourth, for the four-dimensional critical gravity with the Lagrangian
(\ref{4DCriG}) ($c_1=-(2\Lambda)^{-1}$), which is equivalent to
$\sqrt{-g}\big[L_{EH}-1/2L_W(D=4)\big]$, with the help of
Eq. (\ref{KtrQuaC}), $K^{tr}$ given by Eq. (\ref{Ktrcompts}) is
presented by
\begin{equation}
K^{tr}_{4DCG}=\left(K^{tr}_{gr}-\frac{1}{2}K^{tr}_{4DWG}\right)_{D=4}
= K^{tr}_{gr}(D=4)
-\frac{c_1}{2B^3}K^{tr}_{qc}
\, . \label{KtrCriG}
\end{equation}
Particularly, when the metric tensor is the
solution of the vacuum Einstein field equations
$R_{\mu\nu}=\Lambda g_{\mu\nu}$, calculations
show that the potential $K^{tr}_{4DCG}$
disappears. As a matter of fact, this can
be seen straightforwardly from the general
expression for the potential
$K^{\mu\nu}_{4DCG}$ given by Eq. (\ref{Kmn4DCG}).

In the remainder of this section, according to
$K^{tr}$, together with its concrete expressions
in various specifical quadratic-curvature
gravities, we shall give the mass formulations
for the static, spherically symmetric and
asymptotically AdS spacetimes within such
gravity theories. With the metric ansatz
(\ref{SSlineele}), by means of plugging
Eq. (\ref{Ktrcompts}) in Eq. (\ref{Qdef}),
the general formulation for the mass $M$ of
these spacetimes is defined through
\begin{equation}
M = \frac{1}{8\pi}\int_{r=\infty}
\sqrt{h}B^{D-2}K^{tr}d^{D-2}y
\, , \label{MofSSBH}
\end{equation}
where $h=det(h_{ij})$ is the determinant of the
codimension-2 metric $h_{ij}$. As a special case of
Eq. (\ref{MofSSBH}), for Einstein gravity,
Eq. (\ref{Kgrtr}) sends the mass $M$ into the form
\bea
M_{gr} &=&\frac{V_{D-2}}{8\pi}
\lim_{r\rightarrow\infty}\frac{K^{tr}_{gr}}{B^{2-D}}
\nonumber  \\
&=& \frac{V_{D-2}}{32\pi(D-3)\hat{\Lambda}}
\lim_{r\rightarrow\infty}
\frac{\big(2\hat{\Lambda}+f^{\prime\prime}\big)
f^{\prime}}{B^{2-D}}
\, . \label{MofGR}
\eea
In Eq. (\ref{MofGR}),
$V_{D-2}=\int\sqrt{h}d^{D-2}y
=2\pi^{(D-1)/2}/\Gamma\big((D-1)/2\big)$ is the
volume of the $(D-2)$-dimensional unit sphere.
In order to get the second equality, the equation
$f^{\prime\prime}+2\hat{\Lambda}
=(D-2)(D-3)\big(\hat{\Lambda}B^2
+fB^{\prime}B^{\prime}-1\big)/B^2$
derived from the vacuum Einstein field equations
$R^\nu_\mu=(D-1)\hat{\Lambda}\delta^\nu_\mu$ has
been used. As an application of Eq. (\ref{MofGR}),
it can be utilized to evaluate the mass of the
$D$-dimensional Schwarzschild-AdS black holes,
in the line element of which $f(r)$ and $B(r)$
are given by
\be
f_{SAdS}=1-\frac{2m}{r^{D-3}}-\hat{\Lambda}r^2\, ,  \qquad
B_{SAdS}=r
\, , \label{SchAdSMetric}
\ee
respectively. Substituting Eq. (\ref{SchAdSMetric}) into
Eq. (\ref{MofGR}) produces the following mass
\be
M_{SAdS}=\frac{m(D-2)V_{D-2}}{8\pi}
\, , \label{MSchAdS}
\ee
coinciding with the standard result in the literature.
Besides, Eq. (\ref{MofGR}) can be also adopted to compute
the mass for four-dimensional spherical MadMax AdS black holes
constructed quite recently in \cite{HKSTMad}.

In the situation for $D$-dimensional Einstein-Gauss-Bonnet gravity,
the potential $K^{tr}$ in Eq. (\ref{MofSSBH}) is substituted by the
one $K^{tr}_{EGB}$, leading to the mass
\bea
M_{EGB} &=&
\frac{\big[1+2c_1(D-3)(D-4)\hat{\Lambda}\big]V_{D-2}}
{8\pi}
\lim_{r\rightarrow\infty}
\frac{K^{tr}_{gr}}{B^{2-D}} \nn \\
&=&\big[1+2c_1(D-3)(D-4)\hat{\Lambda}\big]M_{gr}
\, . \label{MofEGB}
\eea
Here $M_{gr}$ is defined in terms of the first equality in
Eq. (\ref{MofGR}) and the mass $M_{EGB}$ agrees with the one
via the field-theoretic approach \cite{PeEGB,PetNDT,PetPit}.
Obviously, when $c_1=0$ or $D=4$, $M_{EGB}=M_{gr}$, implying that the
inclusion of the Gauss-Bonnet term in four dimensions does not affect
the mass. For instance, let us apply Eq. (\ref{MofEGB}) to
compute the mass of the static spherically symmetric
asymptotically AdS black holes
in $D$-dimensional $(D>4)$ Einstein-Gauss-Bonnet gravity
\cite{BouDes,GBWhe,GBWil,GBBHcai,GBCNO}. For such black holes,
both the functions $B(r)$ and $f(r)$ in their line elements
are read off as $B_{EGB}=r$ and
\be
f_{EGB}=1+\frac{r^2}{2\tilde{c}_1}-\frac{r^2}{2\tilde{c}_1}
\sqrt{1+4\tilde{c}_1\hat{\Lambda}
+\frac{4\tilde{c}_1m}{r^{D-1}}}
\, , \label{fEGBdef}
\ee
respectively, where $\tilde{c}_1=(D-3)(D-4)c_1$ as before.
Eq. (\ref{MofEGB}) gives rise to the mass
\be
\mathcal{M}_{EGB}=\frac{m(D-2)V_{D-2}}{16\pi}
\, . \label{MasofEGB}
\ee

In the case of the four-dimensional Weyl gravity, the
substitution of Eq. (\ref{Ktr4DW}) into Eq. (\ref{MofSSBH}) gives
rise to the mass
\begin{equation}
M_{4DWG} =
\frac{c_1V_{2}}{8\pi}
\lim_{r\rightarrow\infty}
\frac{K^{tr}_{qc}}{B}
\, . \label{Mof4DWG}
\end{equation}
As a consequence of Eq. (\ref{KtrCriG}), the mass $M_{4DCG}$ for the
four-dimensional critical gravity described by the Lagrangian
(\ref{4DCriG}) can be expressed as the linear combination of
$M_{gr}$ and $M_{4DWG}$, that is,
\be
M_{4DCG} =\left(M_{gr}
-\frac{1}{2}M_{4DWG}\right)_{D=4, c_1=-1/(2\Lambda)}
\, . \label{Mof4DCG}
\ee

Eventually, we take into consideration of some other
applications for the mass formulations (\ref{MofEGB}),
(\ref{Mof4DWG}) and (\ref{Mof4DCG}). Firstly, the formulation
(\ref{MofEGB}) is applicable for the definition of the mass for
the Bardeen-type static spherically-symmetric black holes found
in \cite{BAdSEGB,GSMah,HyNam}, as well as the black holes in
five-dimensional Chern-Simons gravity \cite{BriRa}. Secondly,
as what is shown in Appendix \ref{appendA}, Weyl gravity admits
the four-dimensional Schwarzschild-AdS black hole as its exact solution.
Hence, it is allowed to adopt Eq. (\ref{Mof4DWG}) to define its mass, which
coincides with the one via the (off-shell) ADT and covariant phase
space methods \cite{JJPEPJC}. This holds true for the black holes
given in \cite{StatRWeyl,WeySsMK}. Thirdly, the four-dimensional
Schwarzschild-AdS black hole solution can be embedded into the
four-dimensional critical gravity. For such a solution, the potential
$K^{tr}_{4DCG}=0$ results in its vanishing mass, verifying the
conclusion in \cite{LPcritG}, as well as the one via the
Ashtekar-Magnon-Das method \cite{GOKAMD,PaAMD}. Apart
from the applications mentioned above, it deserves a further
investigation to verify wether the formula (\ref{Qdef}) for
conserved charges can be applicable for the so-called
Buchdahl-inspired metrics presented in
\cite{Nguy23,Ngu232,ANg23}, as well as the (charged)
Lifshitz-type black holes with quadratic-curvature
corrections \cite{BLifBJV,CLS09,LPPV12,BHLif}.

%%%%%%%%%%%%%%%%%%%%%%%%%%%%%%%%%%%%%%%%%%%%%%%%%%%%%%%%%%%%%%%%%%%%%%%%
\section{Mass and angular momentum for rotating spacetimes}\label{four}
%%%%%%%%%%%%%%%%%%%%%%%%%%%%%%%%%%%%%%%%%%%%%%%%%%%%%%%%%%%%%%%%%%%%%%%%

In this section, for the completeness of this study, we shall calculate
the mass and angular momentum of the four-dimensional rotating Kerr-AdS
black hole in the framework of the four-dimensional Weyl, critical and
Einstein-Gauss-Bonnet gravities, respectively. What is more,
the mass and angular momentum for the higher-dimensional
generalizations of the Kerr-AdS black hole will be computed in the context
of the quadratic-curvature gravities with $R^2$ and
$R^{\alpha\beta}R_{\alpha\beta}$ terms. In parallel with the analysis for
Kerr-AdS black holes, we shall take into account black strings
in asymptotically AdS spacetimes.

We adopt the line element for the four-dimensional Kerr-AdS black holes
in a non-rotating frame at infinity, taking the following
form in Boyer-Lindquist coordinates $(t,r,\theta,\phi)$
in $(-,+,+,+)$ notation \cite{4DKeAdSsolu}
\bea
ds^2&=& -\frac{\Delta_r}{\Sigma}\left[dt-a\sin^2\theta
\left(\frac{d\phi}{\Xi} - a\ell^2 \frac{dt}{\Xi}\right)\right]^2
+\frac{\Sigma}{\Delta_{r}}dr^2+\frac{\Sigma}{\Delta_{\theta}}
d\theta^2\nonumber \\
&&+\frac{\Delta_{\theta}\sin^2\theta}{\Sigma}\left[adt-(r^2+a^2)
\left(\frac{d\phi}{\Xi} - a\ell^2\frac{dt}{\Xi}\right)\right]^2
\, . \label{4DKSAdS}
\eea
In Eq. (\ref{4DKSAdS}), the functions $\Delta_{\theta}$, $\Sigma$,
and $\Delta_{r}$ are given by
\begin{align}
\Delta_{\theta}&=1-a^2\ell^2\cos^2\theta \, , \nonumber \\
\Sigma&=r^2+a^2\cos^2\theta \, , \nonumber \\
\Delta_{r}&=(r^2+a^2)(1+\ell^2r^2)-2mr
\, , \label{F4DKSAdS}
\end{align}
respectively. The constant $\ell$, whose inverse
$\ell^{-1}$ stands for the radius of curvature
for the maximally symmetric AdS spaces,
is read off as
$\ell^2=-\hat{\Lambda}_{gr}(D=4)=-\Lambda/3$.
The constant parameter $\Xi = 1-\ell^2 a^2$,
and $(m,a)$ are integration constants related
to the mass and angular momentum respectively.
Due to the fact that the four-dimensional Kerr-AdS
black hole is the solution of the vacuum
Einstein field equation
$R^\nu_\mu=-3\ell^2\delta^\nu_\mu$, the field
equations given by Eq. (\ref{MotionEq2}) support 
that such a black hole is also the exact solution 
of the quadratic gravity characterized by the Lagrangian
\be
\sqrt{-g}L^{(4D)}_{Ric}=\sqrt{-g}\left(R+6\ell^2+c_1 R^2
+c_2R^{\alpha\beta}R_{\alpha\beta}\right)
\, ,  \label{4DLRic}
\ee
where both the coupling constants $c_1$ and $c_2$ are
allowed to be arbitrary in form.

Substituting $\hat{\Lambda}=-\ell^2$ into the 2-form
potential $K_{Ric}^{\mu\nu}$ given by Eq. (\ref{KRic}),
we acquire the potential $K_{KAdS}^{\mu\nu}$ to
define the conserved charges of the four-dimensional
Kerr-AdS black hole (\ref{4DKSAdS}) corrected by
quadratic curvature terms, being of the form
\bea
K_{KAdS}^{\mu\nu}&=&\big[1-6(4c_1+c_2)\ell^2\big]
K_{gr}^{\mu\nu}\big(D=4,\hat{\Lambda}=-\ell^2\big)
\nonumber \\
&=&\big[1-6(4c_1+c_2)\ell^2\big]
\left(\nabla^{[\mu}\xi^{\nu]}+\frac{1}{2\ell^2}
R^{\mu\nu}_{\rho\sigma}\nabla^\rho\xi^\sigma\right)
\, , \label{K4DAdS}
\eea
where the potential $K_{gr}^{\mu\nu}$ for general relativity is
presented by Eq. (\ref{Kgrgen}). Eq. (\ref{K4DAdS}) demonstrates that
the conserved charges in the framework of quadratic gravity
described by the Lagrangian (\ref{4DLRic}) are proportional to
the ones in Einstein gravity. As a consequence, by virtue of the mass
$M_{KAdS}=m/\Xi^2$ and the angular momentum $J_{KAdS}=ma/\Xi^2$
for the four-dimensional Kerr-AdS black hole in general relativity
\cite{PZL,GibPP,BarnichC,DerKather,GullTek,OleKou}, we directly
present its mass $M^{(4DQG)}_{KAdS}$ and angular momentum
$J^{(4DQG)}_{KAdS}$ in the quadratic gravity frame as
\be
M^{(4DQG)}_{KAdS}=\frac{\big[1-6(4c_1+c_2)\ell^2\big]m}{\Xi^2}
\, , \qquad
J^{(4DQG)}_{KAdS}=\frac{\big[1-6(4c_1+c_2)\ell^2\big]ma}{\Xi^2}
\, .\label{MJKin4DQG}
\ee
Particularly, as is shown in Appendix \ref{appendA}, the 
four-dimensional Kerr-AdS black hole is also the solution 
of four-dimensional Weyl gravity ($c_2=-6c_1$) described by
the Lagrangian (\ref{WeylLag}) that does not incorporate 
the Einstein-Hilbert one $\sqrt{-g}(R-2\Lambda)$. Thus, neglecting all the 
contributions from the Einstein-Hilbert Lagrangian, that is,
throwing away $m/\Xi^2$ in $M^{(4DQG)}_{KAdS}$ and $ma/\Xi^2$ 
in $J^{(4DQG)}_{KAdS}$, one obtains the mass $M_{KAdS}^{(4DWG)}$ 
and the angular momentum $J_{KAdS}^{(4DWG)}$ of the 
Kerr-AdS black hole in the framework of Weyl gravity, read 
off as
\begin{equation}
M_{KAdS}^{(4DWG)}=\frac{12c_1\ell^2m}{\Xi^2} \, , \qquad
J_{KAdS}^{(4DWG)}=\frac{12c_1\ell^2ma}{\Xi^2}
\, , \label{MJKin4W}
\end{equation}
respectively. In fact, the mass and angular momentum given 
by the above equation are in agreement with those 
for the four-dimensional charged rotating black hole in 
Weyl gravity \cite{JJPEPJC,LLCRoSW}, which turns into 
the Kerr-AdS black hole in the absence of the charge 
parameter. This is attributed to the fact that the U(1) 
gauge field falls off fast at infinity so that it makes 
no contribution to the conserved charges.

In the case for the theory of critical gravity with the
Lagrangian (\ref{4DCriG}), where $c_1=1/(6\ell^2)$
and $c_2=-3c_1$, the factor $1-6(4c_1+c_2)\ell^2=0$.
Thus, both the mass and angular momentum for
the four-dimensional Kerr-AdS spacetimes vanish.
This holds true for four-dimensional critical
gravity described by the Lagrangian (\ref{4DCriG2}).
To illustrate this, we substitute
$c_1=3\alpha$, $c_2=-12\alpha$ and $c_3=6\alpha$ into
Eq. (\ref{Kmunu}) to obtain the potential
\be
\tilde{K}^{\mu\nu}_{4DCG}=
-6\alpha\Big(R\nabla^{[\mu}\xi^{\nu]}
+4R^{[\mu}_\rho\nabla^{\nu]}\xi^{\rho}
+4\xi^\rho\nabla^{[\mu}R^{\nu]}_\rho\Big)
\, . \label{Kmmn4DCG2}
\ee
Apparently, for the solutions satisfying
$R_{\mu\nu}=3\hat{\Lambda}g_{\mu\nu}$,
the potential $\tilde{K}^{\mu\nu}_{4DCG}=0$, resulting in
identically vanishing conserved charges, which holds true
for the conserved charges defined in terms of the AMD
method \cite{GOKAMD,PaAMD}. This supports the observation
on both the energy and the angular momentum of the
four-dimensional Kerr-AdS black holes in \cite{EdNaCG}.
What is more, in the case for four-dimensional
Einstein-Gauss-Bonnet gravity,
the quadratic-curvature Gauss-Bonnet term makes no contribution
to the mass and the angular momentum of the Kerr-AdS black hole,
arising from that the coupling constants $c_1$ and $c_2$ are
constrained by $4c_1+c_2=0$ in such a gravity theory.
This is also supported by the covariant phase space and the
(off-shell) ADT methods. However, it was demonstrated that
the Gauss-Bonnet term is able to result in a correction to the
Bekenstein-Hawking entropy in \cite{CDGK,JaKMy,PariEntro,CgaPariEnt}.

Without loss of generality, we turn our attention towards the
case for the higher-dimensional Kerr-AdS black holes embedded into
the theory of quadratic-curvature gravity. As is
demonstrated in Appendix \ref{appendB}, the solutions satisfying
$R_{\mu\nu}=(D-1)\hat{\Lambda}g_{\mu\nu}$ ($D>4$)
are also the ones corresponding to the following Lagrangian
\begin{equation}
\sqrt{-g}L_2=\sqrt{-g}\left(R-2\Lambda+c_1 R^2
-Dc_1R^{\alpha\beta}R_{\alpha\beta}
+\frac{(D-2)(\hat{\Lambda}_{gr}-\hat{\Lambda})}{(D-1)(D-4)\hat{\Lambda}^2}
R^{\alpha\beta}R_{\alpha\beta}\right)
\, . \label{Lc1c2}
\end{equation}
Here all three constant parameters $\Lambda$, $c_1$ and
$\hat{\Lambda}$ are allowed to be very general,
which guarantee the constraint (\ref{HatLamcons}) to all
the constant parameters to hold identically. For such solutions,
the 2-form potential $K_{Ric}^{\mu\nu}$ in Eq. (\ref{KRic}),
adopted to define their conserved charges, turns into
\begin{equation}
K_{Ric}^{\mu\nu}\rightarrow
\frac{2(D-2)\hat{\Lambda}_{gr}-D\hat{\Lambda}}{(D-4)\hat{\Lambda}}
K_{gr}^{\mu\nu}
\, . \label{KRicConst}
\end{equation}
In light of the above equation, one sees that
the potential associated with the
Lagrangian (\ref{Lc1c2}) is proportional to
the one for Einstein gravity and  it is
irrelevant to the coupling constant $c_1$.
This is attributed to the fact that the
potential corresponding to the
$c_1 \sqrt{-g}\big(R^2-DR^{\alpha\beta}
R_{\alpha\beta}\big)$
part in the Lagrangian (\ref{Lc1c2})
vanishes identically according to
Eq. (\ref{KmmDDCG2}).

As is evident, the $D$-dimensional Kerr-AdS
black hole \cite{GLuPP1,GLuPP2},
which obeys the field equation
$R_{\mu\nu}=-(D-1)\ell^2 g_{\mu\nu}$
derived from the Einstein-Hilbert Lagrangian
$\mathcal{L}_{EH}=\sqrt{-g}[R+(D-1)(D-2)\ell^2]$,
is an exact solution for the
Lagrangian (\ref{Lc1c2}) with $\hat{\Lambda}=-\ell^2$.
Such a solution has $n=(D-\varepsilon-1)/2$
($\varepsilon=1$ for $D$ even and $\varepsilon=0$
for $D$ odd) independent rotations characterized
by $n$ parameters $a_i$ $(1\leq i\leq n)$ in $n$
orthogonal 2-planes with $2\pi$-periodic azimuthal
angles $\phi_i$. In the coordinate system
$\{t,r,\mu_1,\cdot\cdot\cdot,\mu_{n+\varepsilon-1},
\phi_1,\cdot\cdot\cdot,\phi_n\}$,
the line element for the $D$-dimensional Kerr-AdS black
hole is read off as \cite{GLuPP1,GLuPP2}
\bea
ds^2&=&-HWdt^2+\frac{r^2UV}{H(V-2m)}dr^2
+\sum_{i=1}^{n+\varepsilon}(r^2+a_i^2)\frac{d\mu_i^2}{\Xi_i}
-\frac{\ell^2}{HW}
\left(\sum_{i=1}^{n+\varepsilon}
\frac{r^2+a_i^2}{\Xi_i}\mu_id\mu_i\right)^2
\nonumber \\
&&+\frac{2mH}{r^2UV}
\left(Wdt-\sum_{i=1}^{n}a_i\mu_i^2\frac{d\phi_i}{\Xi_i}\right)^2
+\sum_{i=1}^{n}\mu_i^2(r^2+a_i^2)\frac{d\phi_i^2}{\Xi_i}
\, , \label{MeDKAdS}
\eea
where the four functions $(H,U,V,W)$ are
presented respectively by
\begin{align}
H&=1+\ell^2r^2\, , \quad
U=\sum_{i=1}^{n+\varepsilon}\frac{\mu_i^2}{r^2+a_i^2}
\, , \nonumber \\
V&=r^{\varepsilon-2}H\prod_{i=1}^n(r^2+a_i^2)
\, , \quad
W=\sum_{i=1}^{n+\varepsilon}\frac{\mu_i^2}{\Xi_i}
\, . \label{HWUV}
\end{align}
In Eqs. (\ref{MeDKAdS}) and (\ref{HWUV}), $m$ denotes an integral
constant related to the mass, and the constant parameters
$\Xi_i$ $(1\leq i\leq n)$ are associated with the rotation
parameters $a_i$ $(1\leq i\leq n)$ through $\Xi_i=1-a_i^2\ell^2$,
while $\Xi_{n+1}=1$, arising from that $a_{n+1}=0$ for even $D$.
The $\mu_i$ variables are constrained by
$\sum_{i=1}^{n+\varepsilon}\mu_i^2=1$. By making use of the potential
(\ref{KRicConst}) to compute the mass $M_{qc}$ and the angular momenta
$J^{(i)}_{qc}$($i=1,\cdot\cdot\cdot,n$) for the higher-dimensional
Kerr AdS black holes (\ref{MeDKAdS}) corrected by quadratic curvature
terms, we acquire the results that are proportional to the ones
\cite{PZL,GibPP,BarnichC,DerKather,GullTek,OleKou} for their counterparts
in the framework of general relativity by the factor
$-[2(D-2)\hat{\Lambda}_{gr}+D\ell^2]/[(D-4)\ell^2]$, that is,
\bea
M_{qc}&=&-\frac{\big[2(D-2)\hat{\Lambda}_{gr}+D\ell^2\big]V_{D-2}}
{4\pi(D-4)\ell^2}\frac{m}{\prod_{j=1}^n\Xi_j}
\left(\sum_{i=1}^{n}\frac{1}{\Xi_i}-\frac{1-\varepsilon}{2}\right)
\, , \nonumber \\
J^{(i)}_{qc}&=&-\frac{\big[2(D-2)\hat{\Lambda}_{gr}+D\ell^2\big]V_{D-2}}
{4\pi(D-4)\ell^2}
\frac{ma_i}{\Xi_i\prod_{j=1}^n\Xi_j}
\, . \label{MassAninDdi}
\eea
As it is mentioned in Appendix \ref{appendC}, the above
conserved charges satisfy both the differential
and integral forms for the first law of thermodynamics of black holes.
Particularly, when $\Lambda=-D(D-1)\ell^2/4$ or $\hat{\Lambda}=-\ell^2$,
yielding $\hat{\Lambda}_{gr}=-D\ell^2/[2(D-2)]$, the $D$-dimensional
Kerr-AdS black hole (\ref{MeDKAdS}) is also the exact solution for
$D$-dimensional critical gravity described by the Lagrangian
(\ref {LCriGinD}). In such a case, Eq. (\ref{MassAninDdi}) indicates that
all the mass and the angular momenta of this black hole vanish identically.
In the absence of the cosmological constant, it is worth mentioning that
the higher-derivative corrections to the conserved quantities of
four-dimensional Kerr black holes and static spherically-symmetric
black holes in arbitrary dimensions were taken into account
by virtue of the Euclidean action in \cite{ReaSan19} and \cite{MaPL23}
respectively.

Moreover, we follow the similar procedure to take into consideration of
the case for black strings in asymptotically AdS spacetimes.
It can be verified that the four-dimensional rotating
black strings in Einstein gravity given by the work \cite{LemBstr}
and their higher-dimensional generalization found in \cite{AwaBstr} satisfy
respectively the field equations derived from the Lagrangian (\ref{4DLRic})
with $\ell=1/l$ and the Lagrangian (\ref{Lc1c2}) with $\hat{\Lambda}=-1/l^2$,
where the constant $l$ denotes the radius of AdS spaces
in \cite{LemBstr,AwaBstr}. By analogy with the above case for
the Kerr-AdS black holes, we adopt the 2-form potential $K_{Ric}^{\mu\nu}$
in Eq. (\ref{KRic}) to define the mass and angular momentum for these
black strings. The results for the four-dimensional black strings
and their higher-dimensional generalizations
in the framework of the quadratic-curvature gravity are proportional
to the ones for their counterparts in the context of
general relativity (see the works \cite{LemBstr,AwaBstr,PWmanm} for
the mass and angular momentum of such black strings) by the factors
$1-6(4c_1+c_2)/l^2$ and
$-[2(D-2)l^2\hat{\Lambda}_{gr}+D]/(D-4)$,
respectively.

%%%%%%%%%%%%%%%%%%%%%%%%%%%%%%%%%%%%%%%%%%%%%%%%%%%%%%%%%%%%%%%%%%%%%%%%
\section{Summary}\label{five}
%%%%%%%%%%%%%%%%%%%%%%%%%%%%%%%%%%%%%%%%%%%%%%%%%%%%%%%%%%%%%%%%%%%%%%%%

In this paper, we explored the conserved charges of
asymptotically AdS spacetimes in the context of
quadratic-curvature gravities described by the generic
Lagrangian (\ref{QuaCLag}), which consists of
the Ricci scalar $R$, the cosmological constant $\Lambda$,
as well as the quadratic curvature terms $R^2$,
$R^{\mu\nu}R_{\mu\nu}$ and
$R^{\mu\nu\rho\sigma}R_{\mu\nu\rho\sigma}$. In order to
achieve this, we first analyzed the structure of the
Lagrangian (\ref{QuaCLag}) and derived the expression
for the equations of motion via the variation of this
Lagrangian. Next, through the linear combination of the
rank-4 tensors $P^{\mu\nu}_{\rho\sigma}$ and
$P^{\mu\nu}_{(ref)\rho\sigma}$, we defined the rank-4
tensor $\mathcal{P}^{\mu\nu}_{\rho\sigma}$ given
by Eq. (\ref{CalPdef}), which inherits the index symmetries
of the Riemann curvature tensor and disappears on the
AdS spacetimes. By the aid of
$\mathcal{P}^{\mu\nu}_{\rho\sigma}$, a 2-form potential
$K^{\mu\nu}$ associated to the Killing vector field
was proposed in Eq. (\ref{Kmunu0}), which
resembles the Noether potential $K^{\mu\nu}_R$
in Eq. (\ref{KRmunu}). To demonstrate that $K^{\mu\nu}$
is suitable for the definition of conserved charges
of asymptotically AdS spacetimes, we compared it with
the results via other methods, such as the covariant
phase space approach, the (off-shell) ADT formalism,
the generalized Komar integral and the field-theoretic
method. Furthermore, in terms of the potential $K^{\mu\nu}$,
the formula (\ref{Qdef}) for conserved charges was
presented. Finally, as applications, we derived the mass
formula (\ref{MofSSBH}) for static and spherically
symmetric spacetimes in four dimensions and above.
Besides, the formula (\ref{Qdef})
was applied to compute the mass and the angular momentum
of the four(higher)-dimensional Kerr-AdS black holes
and black strings embedded into quadratic-curvature
gravities, which are proportional to the ones
in Einstein gravity. All the results reveal that the potential
$K^{\mu\nu}$ can successfully give rise to a simple
formula for the conserved charges of asymptotically AdS
spacetimes in the theories of quadratic-curvature gravity.

Particularly, we stressed on the definition of conserved quantities
for Weyl gravity, Einstein-Gauss-Bonnet gravity and critical
gravity. For these theories, the potentials adopted
to define conserved charges were given respectively by
Eqs. (\ref{KmunuW}), (\ref{KmnEGB}) and (\ref{Kmn4DCG}).
What is more, for the asymptotically AdS spacetimes
satisfying the vacuum Einstein field equations, the potential
$K_{Ric}^{\mu\nu}$ given by Eq. (\ref{KRic}) takes a simple
form, which is proportional to the one for Einstein gravity.

\section*{Acknowledgments}

This work was supported by the Natural Science Foundation of
China under Grant Nos. 11865006. It was also partially supported
by the Technology Department of Guizhou province Fund
under Grant Nos. [2018]5769.

%%%%%%%%%%%%%%%%%%%%%%%%%%%%%%%%%%%%%%%%%%%%%%%%%%%%%%%%%%%
\appendix
%%%%%%%%%%%%%%%%%%%%%%%%%%%%%%%%%%%%%%%%%%%%%%%%%%%%%%%%
\section{The properties for the tensor $P^{\mu\nu\rho\sigma}$
and the equations of motion} \label{appendA}
%%%%%%%%%%%%%%%%%%%%%%%%%%%%%%%%%%%%%%%%%%%%%%%%%%%%

We devote this appendix to the properties of the rank-4 tensor
$P^{\mu\nu\rho\sigma}$ and the divergence for the expression of the
equations of motion, together with the field equations associated
with four-dimensional Weyl and critical gravities and
Einstein-Gauss-Bonnet gravity in arbitrary dimensions. Some of the
following results for $P^{\mu\nu\rho\sigma}$ overlap the ones
in \cite{Pady}.

In order to acquire the rank-4 tensor $P^{\mu\nu\rho\sigma}$ given by
Eq. (\ref{PRtensor}), we evaluate the derivative
of the Riemann tensor $ R_{\alpha\beta\gamma\lambda}$ with respect to
$R^{\rho\sigma}_{\mu\nu}$, yielding
\begin{equation}
\frac{\partial R_{\alpha\beta\gamma\lambda}}
{\partial R^{\rho\sigma}_{\mu\nu}}
= \frac{1}{8}\big(g_{\alpha\kappa}g_{\beta\eta}
\delta^{\kappa\eta}_{\rho\sigma}\delta_{\gamma\lambda}^{\mu\nu}
+g_{\gamma\kappa}g_{\lambda\eta}
\delta^{\kappa\eta}_{\rho\sigma}\delta_{\alpha\beta}^{\mu\nu}\big)
\, ,\label{ParRRR}
\end{equation}
whose contraction with the metric tensor further gives rise to
\bea
\frac{\partial R_{\gamma\lambda}}
{\partial R^{\rho\sigma}_{\mu\nu}}
&=&g^{\alpha\beta}\frac{\partial R_{\alpha\gamma\beta\lambda}}
{\partial R^{\rho\sigma}_{\mu\nu}}
= -\frac{1}{4}\big(g_{\gamma[\rho}
\delta^{\mu\nu}_{\sigma]\lambda}+g_{\lambda[\rho}
\delta^{\mu\nu}_{\sigma]\gamma}\big)\, , \nonumber  \\
\frac{\partial R}{\partial R^{\rho\sigma}_{\mu\nu}}
&=&g^{\gamma\lambda}\frac{\partial R_{\gamma\lambda}}
{\partial R^{\rho\sigma}_{\mu\nu}}
= \frac{1}{2}\delta^{\mu\nu}_{\rho\sigma}
\, . \label{ParRRR1}
\eea

According to the definition for the rank-4 tensor
$P^{\mu\nu\rho\sigma}$, it is required to at least
inherit the following algebraic symmetries for
the Riemann tensor $R^{\mu\nu\rho\sigma}$, that is,
\be
P^{\mu\nu\rho\sigma}=P^{\rho\sigma\mu\nu}
=-P^{\nu\mu\rho\sigma}=-P^{\mu\nu\sigma\rho}
\, . \label{PindeSym}
\ee
Particularly, it can be proved that the tensor
$P^{\mu\nu\rho\sigma}$ also fulfills the algebraic
Bianchi-type identities
$P^{\mu[\nu\rho\sigma]}=0=P^{[\mu\nu\rho]\sigma}$ in
the case of quadratic-curvature gravities. Besides,
by merely making use of the properties of the Riemann
tensor and the index symmetries of $P_{\mu\nu\rho\sigma}$
presented by Eq. (\ref{PindeSym}), we have
\bea
2\nabla^{[\rho} \nabla^{\sigma]}P_{\mu\nu\rho\sigma}&=&
-2R_{[\mu}^{~~\lambda\rho\sigma}P_{\nu]\lambda\rho\sigma}
+R_{\rho}^{~~\lambda\rho\sigma}P_{\mu\nu\lambda\sigma}
+R_{\sigma}^{~~\lambda\rho\sigma}P_{\mu\nu\rho\lambda} \, , \nn \\
&=&-2R_{[\mu}^{~~\lambda\rho\sigma}P_{\nu]\lambda\rho\sigma}
+2R^{\rho\sigma}P_{\mu\nu\rho\sigma}
=-2R_{[\mu}^{~~\lambda\rho\sigma}P_{\nu]\lambda\rho\sigma}
\, ,  \label{DivRP2t}
\eea
together with the following identity
\bea
2\nabla^{[\rho} \nabla^{\sigma]}P_{\rho\mu\nu\sigma}
&=&R_{[\mu}^{~~\lambda\rho\sigma}P_{\nu]\sigma\rho\lambda}
+R_{[\mu}^{~~\lambda\rho\sigma}P_{\nu]\rho\lambda\sigma}
\nn \\
&=&R_{[\mu}^{~~\lambda\rho\sigma}P_{\nu]\lambda\rho\sigma}
\, . \label{IdenP2}
\eea
In order to gain the last equality in Eq. (\ref{IdenP2}),
we have used the Bianchi identity
$R_{\mu[\lambda\rho\sigma]}=0$.
As a consequence of Eqs. (\ref{DivRP2t}) and (\ref{IdenP2}),
unlike in \cite{Pady}, without the requirement that
$R_{[\mu}^{~~\lambda\rho\sigma}P_{\nu]\lambda\rho\sigma}=0$
and $P^{\rho[\mu\nu\sigma]}=0$, their
combination brings about another identity
\be
\nabla^\rho\nabla^\sigma P_{\mu[\nu\rho\sigma]}=0
\, , \label{IdenP3}
\ee
or $\nabla^\rho\nabla^\sigma P_{\mu\nu\rho\sigma}
=-2\nabla^\rho\nabla^\sigma P_{\rho[\mu\nu]\sigma}$. This can
be also obtained by using
$P^{[\mu\nu\rho\sigma]}=P^{\mu[\nu\rho\sigma]}$ and
$\partial_\rho\partial_\sigma
\big(\sqrt{-g}P^{[\mu\nu\rho\sigma]}\big)=0$.
Moreover, by the aid of the equality (\ref{IdenP2}), as well as
the decompositions to the rank-2 tensors
$R_{\mu}^{~~\lambda\rho\sigma}P_{\nu\lambda\rho\sigma}$
and
$\nabla^{\rho} \nabla^{\sigma}P_{\rho\mu\nu\sigma}$, that is,
\bea
R_{\mu}^{~~\lambda\rho\sigma}P_{\nu\lambda\rho\sigma}&=&
R_{(\mu}^{~~\lambda\rho\sigma}P_{\nu)\lambda\rho\sigma}
+R_{[\mu}^{~~\lambda\rho\sigma}P_{\nu]\lambda\rho\sigma} \, , \nn \\
\nabla^{\rho} \nabla^{\sigma}P_{\rho\mu\nu\sigma}&=&
\nabla^{\rho} \nabla^{\sigma}P_{\rho(\mu\nu)\sigma}
+\nabla^{\rho} \nabla^{\sigma}P_{\rho[\mu\nu]\sigma}
\, ,
\eea
we obtain the following identity
\be
R_{\mu}^{~~\lambda\rho\sigma}P_{\nu\lambda\rho\sigma}
-2\nabla^{\rho} \nabla^{\sigma}P_{\rho\mu\nu\sigma}
=R_{(\mu}^{~~\lambda\rho\sigma}P_{\nu)\lambda\rho\sigma}
-2\nabla^{\rho} \nabla^{\sigma}P_{\rho(\mu\nu)\sigma}
\,  \label{Rm2delP}
\ee
under the only requirement that $P_{\mu\nu\rho\sigma}$
exhibits the index symmetries in Eq. (\ref{PindeSym}).
As a result of Eq. (\ref{Rm2delP}), the round brackets in
the expression (\ref{MotionEq}) for the field equations
can be omitted.

In the case where $P_{\mu\nu\rho\sigma}$ is not required to
satisfy $P_{\mu[\nu\rho\sigma]}=0$, we can introduce an
auxiliary rank-4 tensor $\tilde{P}_{\mu\nu\rho\sigma}$
to decompose $P_{\mu\nu\rho\sigma}$ as
\be
P_{\mu\nu\rho\sigma}=\tilde{P}_{\mu\nu\rho\sigma}
+P_{\mu[\nu\rho\sigma]}
\, . \label{tildP}
\ee
Obviously, $\tilde{P}_{\mu\nu\rho\sigma}$ possesses the index
symmetries given by Eq. (\ref{PindeSym}) and satisfies the
Bianchi-type identity $\tilde{P}_{\mu[\nu\rho\sigma]}=0$.
By means of
$R_{\mu}^{~~\lambda\rho\sigma}P_{\nu[\lambda\rho\sigma]}
=R_{\mu}^{~~[\lambda\rho\sigma]}P_{\nu[\lambda\rho\sigma]}=0$
and Eq. (\ref{IdenP3}), we have
\be
R_{\mu}^{~~\lambda\rho\sigma}P_{\nu\lambda\rho\sigma}
-2\nabla^{\rho} \nabla^{\sigma}P_{\rho\mu\nu\sigma}
=R_{\mu}^{~~\lambda\rho\sigma}\tilde{P}_{\nu\lambda\rho\sigma}
-2\nabla^{\rho} \nabla^{\sigma}\tilde{P}_{\rho\mu\nu\sigma}
\, . \label{Rm2tildP}
\ee
As a matter of fact, the expression (\ref{MotionEq}) for
the field equations is applicable to the more general
Lagrangian depending on the metric and the Riemann tensor.
In terms of Eq. (\ref{Rm2tildP}), we conclude that the
equations of motion associated to such types of Lagrangian
are irrelevant to $P_{\mu[\nu\rho\sigma]}$. This holds
true for the surface term $\Theta^\mu$ arising from that
$P^{\mu[\nu\rho\sigma]}\nabla_{[\sigma}\delta g_{\rho\nu]}=0$ and
$\delta g_{[\nu\rho} \nabla_{\sigma]} P^{\mu[\nu\rho\sigma]}=0$.
Thus, the variation of the Lagrangian that is the functional of
the metric and the Riemann tensor is alternatively given by
\be
\delta\big(\sqrt{-g}L\big)=\sqrt{-g}\tilde{E}_{\mu\nu}
\delta g^{\mu\nu}+\sqrt{-g}\nabla_\mu \tilde{\Theta}^\mu
\, , \label{LagvaTild}
\ee
where $\tilde{E}_{\mu\nu}=E_{\mu\nu}\big(P\rightarrow\tilde{P}\big)$
and $\tilde{\Theta}^{\mu}=\Theta^\mu\big(P\rightarrow\tilde{P}\big)$.
According to Eq. (\ref{LagvaTild}), one can follow the conventional
Noether procedure to obtain the Noether
current and potential unrelated to $P_{\mu[\nu\rho\sigma]}$,
as well as the Iyer-Wald potential. Hereto we stress that
all the above results hold true for an arbitrary rank-4 tensor
$P_{\mu\nu\rho\sigma}$ only armed with the index symmetries
given by Eq. (\ref{PindeSym}). If we further make use of the
result $R_{\mu}^{~~\lambda\rho\sigma}P_{\nu\lambda\rho\sigma}
=R_{\nu}^{~~\lambda\rho\sigma}P_{\mu\lambda\rho\sigma}$, which
was explicitly proved under the conditions that the Lagrangian
preserves diffeomorphism invariance and the $P$-tensor
exhibits the algebraic symmetries presented by Eq. (\ref{PindeSym})
in \cite{Pady}, Eqs. (\ref{DivRP2t}) and (\ref{IdenP2})
turn into
\be
\nabla^{\rho} \nabla^{\sigma}P_{\mu\nu\rho\sigma}=0
\, , \qquad
\nabla^{\rho} \nabla^{\sigma}P_{\rho[\mu\nu]\sigma}=0
\, . \label{Cov2P}
\ee

The results in Eq. (\ref{Cov2P}) will be verified below
in the framework of quadratic-curvature gravities.
Specifically, the divergence of $P^{\mu\nu\rho\sigma}$
takes the form
\be
\nabla^\rho P^{\mu\nu}_{\rho\sigma} = \frac{1}{4}(4c_1+c_2)
\delta^{\mu\nu}_{\rho\sigma}
\nabla^\rho R+(c_2+4c_3)\nabla^{[\mu}R^{\nu]}_\sigma
\, .  \label{divPR}
\ee
Calculations on the divergence of $\nabla^\rho P^{\mu\nu}_{\rho\sigma}$
give rise to
\bea
\nabla_\rho\nabla^\sigma P^{\mu\rho}_{\sigma\nu}
&=&-\frac{1}{4}(4c_1+c_2)\delta^{\mu}_{\nu}\Box R
-\frac{1}{2}(c_2+4c_3)\Box R^{\mu}_{\nu}
\nonumber  \\
&&+\frac{1}{2}(c_2+4c_3)\big(R^{\mu}_{\lambda}R^\lambda_\nu-
R^{\mu}_{~~\rho\nu\sigma}R^{\rho\sigma}\big)
\nonumber  \\
&&+\frac{1}{2}(2c_1+c_2+2c_3)\nabla^\mu\nabla_\nu R
\, ,  \label{TwodivP}
\eea
as well as $\nabla^\rho \nabla^\sigma P^{\mu\nu}_{\rho\sigma}
=\nabla^{[\rho} \nabla^{\sigma]}P^{\mu\nu}_{\rho\sigma}=0$.
The contraction between the indices $\mu$ and $\nu$ in
$\nabla_\rho\nabla^\sigma P^{\mu\rho}_{\sigma\nu}$ yields
the scalar
\be
g^{\mu\nu}\nabla^\rho\nabla^\sigma P_{\mu\rho\sigma\nu}
=-\frac{1}{4}[4(D-1)c_1+Dc_2+4c_3]\Box R
\, . \label{Con2delP}
\ee
In terms of Eq. (\ref{DivRP2t}),
the equality $\nabla^\rho\nabla^\sigma P_{\mu\nu\rho\sigma}=0$
renders us to arrive at
$R_{[\mu}^{~~\lambda\rho\sigma}P_{\nu]\lambda\rho\sigma}=0$, or
\be
R_{\mu}^{~~\lambda\rho\sigma}P_{\nu\lambda\rho\sigma}
=R_{\nu}^{~~\lambda\rho\sigma}P_{\mu\lambda\rho\sigma}
\, . \label{RPsym}
\ee
Due to Eq. (\ref{IdenP2}) or (\ref{IdenP3}), Eq. (\ref{RPsym})
further results in
\begin{equation}
\nabla^\rho\nabla^\sigma P_{\rho\mu\nu\sigma}
=\nabla^\rho\nabla^\sigma P_{\rho\nu\mu\sigma}
\, , \label{DivP2sym}
\end{equation}
or $\nabla_\rho\nabla_\sigma P^{\rho[\mu\nu]\sigma}
=\nabla_{[\rho}\nabla_{\sigma]} P^{\rho\mu\nu\sigma}
=0$.
Therefore, we have the conclusion that
$\nabla^\rho\nabla^\sigma P_{\mu\nu\rho\sigma}=0$
is the necessary and sufficient condition for the result that each
of the rank-2 tensors $\nabla^\rho\nabla^\sigma P_{\rho\mu\nu\sigma}$
and $R_{\mu}^{~~\lambda\rho\sigma}P_{\nu\lambda\rho\sigma}$
is symmetric with respect to both the indices $\mu$ and $\nu$.

With the help of the tensor $P^{\mu\nu\rho\sigma}$, the Lagrangian density
$L$ can be reexpressed as
\be
2L=R-4\Lambda+R_{\mu\nu}^{\rho\sigma}P^{\mu\nu}_{\rho\sigma}
\, . \label{Laltexpre}
\ee
When the metric satisfies the field equations $E_{\mu\nu}=0$,
by the aid of Eq. (\ref{Con2delP}), $L$ is required to fulfill
the following on-shell condition
\be
(D-4)L=[4(D-1)c_1+Dc_2+4c_3]\Box R-2R+8\Lambda
\, , \label{Lonshel}
\ee
or equivalently,
\be
(D-4)R_{\mu\nu}^{\rho\sigma}P^{\mu\nu}_{\rho\sigma}
=[8(D-1)c_1+2Dc_2+8c_3]\Box R-DR+4D\Lambda
\, . \label{RPonshel}
\ee
In particular, when the dimension of spacetimes $D=4$, the
constraint (\ref{Lonshel}) or (\ref{RPonshel}) for all the
quadratic-curvature gravities is simplified as
\be
2[3c_1+c_2+c_3]\Box R=R-4\Lambda
\, . \label{4DLonshel}
\ee
For four-dimensional Weyl gravity coupled with the Einstein-Hilbert
Lagrangian and critical gravity described by the Lagrangian
(\ref{4DCriG}), one obtains $R=4\Lambda$.

In terms of the above properties of the tensor $P^{\mu\nu\rho\sigma}$,
the divergence of the expression for the equations of motion $E_{\mu\nu}$
is read off as
\bea
\nabla_\nu E_{\mu}^\nu&=&\frac{1}{2}
P_{\rho\sigma}^{\lambda\nu}
\nabla_{\lambda}R_{\mu\nu}^{\rho\sigma}
 -\frac{1}{4}R_{\alpha\beta}^{\rho\sigma}
\nabla_{\mu}P_{\rho\sigma}^{\alpha\beta}
-\frac{1}{4}\nabla_\mu R
 \nonumber  \\
&=&\frac{1}{4}\big[(2c_1R+1)\big(2\nabla_\nu R_{\mu}^\nu
-\nabla_\mu R\big)
\nonumber  \\
&&+2c_2 R_{\rho}^\nu\big(\delta_{\mu\nu}^{\gamma\lambda}
\nabla_{\lambda} R_{\gamma}^\rho
-\nabla_\sigma R_{\mu\nu}^{\rho\sigma}\big)
\nonumber  \\
&&+2c_3R^{\gamma\lambda}_{\rho\sigma}
\big(2\nabla_\gamma R_{\mu\lambda}^{\rho\sigma}
-\nabla_\mu R_{\gamma\lambda}^{\rho\sigma} \big)
\big]
 \nonumber  \\
&\equiv& 0
\, .  \label{DivEmn}
\eea
In order to obtain the last identity in the above equation, the
Bianchi identity $\nabla_{[\mu} R_{\gamma\lambda]}^{\rho\sigma}=0$
has been used. Eq. (\ref{DivEmn}) shows that the expression for
the equations of motion is conserved.

In the remainder of this appendix, to demonstrate some typical cases of the
general expression (\ref{MotionEq2}) for the field equations, we
take into account the equations of motion for the four-dimensional
Weyl and critical gravities, as well as the one for the Einstein-Gauss-Bonnet
gravity in arbitrary dimensions. It can be proven
that the Gauss-Bonnet term (\ref{GBterm}) fulfills identically
\bea
g_{\mu\nu}L_{GB} &=&
4RR_{\mu\nu}-8R_{\mu\rho\nu\sigma}R^{\rho\sigma}
-8R_{\mu\lambda}R^\lambda_\nu
\nn  \\
&&+4R_{\mu}^{~~\lambda\rho\sigma}R_{\nu\lambda\rho\sigma}
+\frac{1}{4}g_{\nu\zeta}
\delta^{\zeta\gamma\lambda\kappa\eta}_{\mu\alpha\beta\rho\sigma}
R^{\alpha\beta}_{\gamma\lambda}R^{\rho\sigma}_{\kappa\eta}
\, .  \label{Curvident}
\eea
Attributed to the fact that
 $\delta^{\zeta\gamma\lambda\kappa\eta}_{\mu\alpha\beta\rho\sigma}=0$
when $D\leq4$, Eq. (\ref{Curvident}) can be used to simplify the
expression $E_{\mu\nu}$ for field equations in four dimensions.
Doing so yields $E^{(4D)}_{\mu\nu}
=E_{\mu\nu}|_{c_3=0}(c_1,c_2
\rightarrow\hat{c}_1,\hat{c}_2)$, where $\hat{c}_1=c_1-c_3$
and $\hat{c}_2=c_2+4c_3$, namely,
\bea
E^{(4D)}_{\mu\nu}&=&E^{(gr)}_{\mu\nu}+2\hat{c}_1 RR_{\mu\nu}
+2\hat{c}_2 R_{\mu\rho\nu\sigma}R^{\rho\sigma}
-\frac{1}{2}g_{\mu\nu}\Big(\hat{c}_1 R^2
+\hat{c}_2 R_{\alpha\beta}R^{\alpha\beta}\Big)
 \nn  \\
&&+\frac{1}{2}(4\hat{c}_1+\hat{c}_2)g_{\mu\nu}\Box R
+\hat{c}_2\Box R_{\mu\nu}
-(2\hat{c}_1+\hat{c}_2)\nabla_\mu\nabla_\nu R
\, .  \label{4DMotEq}
\eea
Here $E^{(gr)}_{\mu\nu}$ denotes the expression for
field equations corresponding to the Einstein-Hilbert
Lagrangian
$\sqrt{-g}L_{EH}=\sqrt{-g}(R-2\Lambda)$, and it reads
\be
E^{(gr)}_{\mu\nu}=R_{\mu\nu}-\frac{1}{2}R g_{\mu\nu}
+\Lambda g_{\mu\nu}
\, . \label{EoMofGR}
\ee
From Eq. (\ref{4DMotEq}), one observes that the solution
satisfying $E^{(gr)}_{\mu\nu}=0$ or
$R_{\mu\nu}=\Lambda g_{\mu\nu}$ in four dimensions
must guarantee that $E^{(4D)}_{\mu\nu}=0$. That is
to say, the vacuum solution to four-dimensional
general relativity must be the exact one
for four-dimensional quadratic gravity \cite{PPS17}.
As a result of Eq. (\ref{EoMofGR}), the expression
for the field equations of four-dimensional Weyl
gravity, corresponding to $E^{(4D)}_{\mu\nu}$
with $c_2=-6c_1$ and $c_3=3c_1$,
or with $\hat{c}_1=-2c_1$ and $\hat{c}_2=6c_1$,
is given by
\bea
E^{(4DW)}_{\mu\nu}
&=&c_1\big(R^2-3R_{\rho\sigma}R^{\rho\sigma}
-\Box R \big)g_{\mu\nu}
+6c_1\Box R_{\mu\nu}
\nn  \\
&&+2c_1\big(6R_{\mu\rho\nu\sigma}R^{\rho\sigma}
-2RR_{\mu\nu}
-\nabla_\mu\nabla_\nu R\big)
\nn  \\
&=&-6c_1\big(2\nabla^\rho\nabla^\sigma+R^{\rho\sigma}\big)
C_{\mu\rho\sigma\nu}
\, ,  \label{EqM4DWeyl}
\eea
where Eq. (\ref{Curvident}) has been adopted to
expand the four-dimensional Bach tensor
$(\nabla^\rho\nabla^\sigma
+R^{\rho\sigma}/2)C_{\mu\rho\nu\sigma}$.
According to Eq. (\ref{EqM4DWeyl}), one finds that
$g^{\mu\nu}E^{(4DW)}_{\mu\nu}=0$ and the condition
$R_{\mu\nu}=\lambda g_{\mu\nu}$, where $\lambda$ is an arbitrary
constant, must give rise to $E^{(4DW)}_{\mu\nu}=0$.
This indicates that the four-dimensional vacuum general
relativistic solution is the one for the four-dimensional
Weyl gravity as well.

When $c_1=\beta$, $c_2=\alpha$ and $c_3=0$, Eq. (\ref{MotionEq2}) becomes
the expression for the field equation of the four-dimensional
critical gravity given in \cite{LPcritG}. Specifically, according to the
Lagrangian (\ref{4DCriG}), the expression $E^{(4DCG)}_{\mu\nu}$
for the field equation of the four-dimensional critical gravity
can be related to the one $E^{(4DW)}_{\mu\nu}$ for the four-dimensional
Weyl gravity through
\be
E^{(4DCG)}_{\mu\nu}=E^{(gr)}_{\mu\nu}
+E^{(4DW)}_{\mu\nu}\Big(c_1=\frac{1}{4\Lambda}\Big)
\, .  \label{EqM4DCG}
\ee
It should be pointed out that the higher-dimensional generalization
of Eq. (\ref{EqM4DCG}) can be found in the work \cite{KKSCriG}
(see also Eq. (\ref{LCriGinD}) in Appendix \ref{appendB}).
What is more, with the help of Eq. (\ref{Curvident}),
neglecting the $E^{\nu}_{(gr)\mu}$ part and letting
$c_1=3\alpha$, $c_2=-12\alpha$ and $c_3=6\alpha$ in Eq. (\ref{MotionEq2}),
we get the expression for the field equation of the critical gravity
described by the Lagrangian (\ref{4DCriG2}), being of the form
\bea
\tilde{E}^{(4DCG)}_{\mu\nu}
&=&6\alpha\big(4R_{\mu\rho\nu\sigma}R^{\rho\sigma}-RR_{\mu\nu}
-\nabla_\mu\nabla_\nu R\big)
\nonumber  \\
&&+\frac{3}{2}\alpha\big(R^2-4R_{\rho\sigma}R^{\rho\sigma}\big)g_{\mu\nu}
+12\alpha\Box R_{\mu\nu}
\, .  \label{EqM4DCG2}
\eea
Due to the fact that $g^{\mu\nu}\tilde{E}^{(4DCG)}_{\mu\nu}
=6\alpha\Box R$, one necessary condition for
$\tilde{E}^{(4DCG)}_{\mu\nu}=0$ is $\Box R=0$.
With Eqs. (\ref{EqM4DWeyl}) and (\ref{EqM4DCG}), solutions obeying
$E^{\nu}_{(gr)\mu}(D=4)=0$ must be the ones of the critical gravity
relative to the Lagrangian (\ref{4DCriG}). This can also be seen from
the Lagrangian (\ref{4DCriG}) attributed to the fact that the Gauss-Bonnet
term in four dimensions is a topological surface term, which makes
no contribution to the equations of motion. According to
Eq (\ref{EqM4DCG2}), the critical gravity with the
Lagrangian (\ref{4DCriG2}) allows for the solutions satisfying
$R_{\mu\nu}=\lambda g_{\mu\nu}$. As a result, in light of all the above
in this appendix, one observes that the AdS black hole solution with
cylindrical symmetry found in \cite{SASclin} can be embedded in the
four-dimensional Weyl and critical gravities as well.

In addition, in the case for Einstein-Gauss-Bonnet
gravity ($c_2=-4c_1$, $c_3=c_1$),
Eq. (\ref{MotionEq2}) coincides with the field equation
in \cite{JJPEPJC}, which can be further simplified
as the following form by virtue of Eq. (\ref{Curvident})
\begin{equation}
E^\nu_{(GB)\mu}=E^{\nu}_{(gr)\mu}-\frac{c_1}{8}
\delta^{\nu\gamma\lambda\kappa\eta}_{\mu\alpha\beta\rho\sigma}
R^{\alpha\beta}_{\gamma\lambda}R^{\rho\sigma}_{\kappa\eta}
\, . \label{EmotofEGB}
\end{equation}
In terms of Eq. (\ref{EmotofEGB}), it is easy to see that the
Gauss-Bonnet term is non-dynamical in four dimensions due to
the fact that
$\delta^{\nu\gamma\lambda\kappa\eta}_{\mu\alpha\beta\rho\sigma}\equiv0$
in four dimensions.

%%%%%%%%%%%%%%%%%%%%%%%%%%%%%%%%%%%%%%%%%%%%%%%%%%%%%%%%
\section{Embedding higher-dimensional solutions to
vacuum Einstein field equations into quadratic gravity} \label{appendB}
%%%%%%%%%%%%%%%%%%%%%%%%%%%%%%%%%%%%%%%%%%%%%%%%%%%%
%The solutions satisfying
%$R_{\mu\nu}=(D-1)\hat{\Lambda}g_{\mu\nu}$ in the context of
%Lagrangian (\ref{Lc1c2})

In order to test the definition for conserved quantities
in quadratic gravity described by the Lagrangian
$\sqrt{-g}L$, enough exact solutions of this theory
in various dimensions are desired. Unfortunately,
since the field equations $E_{\mu\nu}=0$ are
non-linear fourth-order partial
differential equations (PDE), solving them to obtain
solutions is extremely difficult. However,
within the context of some special cases
of quadratic gravity or it coupled with matter fields,
such as Einstein-Gauss-Bonnet gravity, Weyl gravity
and critical gravity, some exact static spherically
symmetric solutions were found in the literature.
Particularly, it has been demonstrated in Appendix
\ref{appendA} that all the vacuum solutions
to four-dimensional Einstein gravity automatically
become the ones for four-dimensional quadratic
gravity. Some non-Einstein vacuum solutions to
four-dimensional quadratic gravity were obtained
in \cite{SPPP,PSPP20,PPPS21,PPS17}.
Apart from the aforementioned solutions,
to our knowledge, within the framework of the full
theory of quadratic gravity,
all exact solutions in any dimension found so far
are Kundt spacetimes with constant Ricci scalar
constructed in \cite{GGST11,GST12}, which can be
put into Kerr-Schild form. Due to the complexity
of the field equations, exact rotating solutions
in $D>4$ dimensions are still absent up till now.

In the present appendix, let us focus on the
solutions to the field equations $E_{\mu\nu}=0$ in
Eq. (\ref{MotionEq2}) under the sole constraint
condition that they obey the vacuum Einstein
field equations $R_{\mu\nu}=\lambda g_{\mu\nu}$
in arbitrary dimensions, where $\lambda$ represents
a constant. Such a constraint to them gets rid of
all the fourth-order derivative terms $\Box R$,
$\Box R_{\mu\nu}$ and
$\nabla_{\mu}\nabla_{\nu} R$ in the field
equations. Hence, the
original fourth-order field equations reduce to
the much simpler second-order ones depending only on
the metric and curvature terms. Then the substitution
of $R_{\mu\nu}=\lambda g_{\mu\nu}$ into the
resulting field equations gives rise to
\bea
E^\nu_\mu\big|_{R_{\rho\sigma}=\lambda
g_{\rho\sigma}}&=&-\frac{1}{2}\Big[
(D-4)(Dc_1+c_2)\lambda^2
+(D-2)\lambda-2\Lambda
\Big]\delta^\nu_\mu \nn \\
&&+\frac{c_3}{2}\big(4R^{\rho\sigma}_{\mu\lambda}
R^{\nu\lambda}_{\rho\sigma}
- R^{\rho\sigma}_{\alpha\beta}
R^{\alpha\beta}_{\rho\sigma}\delta^\nu_\mu \big)
=0
\, . \label{EoMundGR}
\eea
The four free parameters $(\Lambda,c_1,c_2,c_3)$
in field equations allow for their appropriate
values to guarantee that Eq. (\ref{EoMundGR})
holds without any further requirement for the
solutions except for the existing constraint
$R_{\mu\nu}=\lambda g_{\mu\nu}$.
(this will be illustrated below by Eqs.
(\ref{MotEqL2}), (\ref{OnePaLag}), (\ref{DDCriG2})
and (\ref{TwoPaLag})). That is to say, like in
the four-dimensional case, it is also possible
to embed the vacuum solutions in higher-dimensional
Einstein gravity into quadratic gravities in higher
dimensions. Under the guidance of the above, we will
demonstrate in detail that the solutions of the
vacuum Einstein equations
$R_{\mu\nu}=(D-1)\hat{\Lambda}g_{\mu\nu}$
(here the dimension of spacetime $D>4$ and the
non-vanishing constant $\hat{\Lambda}$ is allowed
to be arbitrary) belong to the quadratic gravities
with the Lagrangian (\ref{Lc1c2}) as well.

Substituting equations
$R_{\mu\nu}=(D-1)\hat{\Lambda}g_{\mu\nu}$ and
$R=D(D-1)\hat{\Lambda}$ into Eq. (\ref{Lc1c2}), we have
\be
L_2=D(D-1)\Big[(D-1)\big(Dc_1+\tilde{c}_2\big)\hat{\Lambda}^2
+\hat{\Lambda}\Big]-2\Lambda
\, ,  \label{L2forRic}
\ee
where the constant $\tilde{c}_2$ is read off as
\be
\tilde{c}_2=
\frac{(D-2)(\hat{\Lambda}_{gr}
-\hat{\Lambda})}{(D-1)(D-4)\hat{\Lambda}^2}
-Dc_1
\, . \label{Tildc2}
\ee
Under the condition that $R_{\mu\nu}
=(D-1)\hat{\Lambda}g_{\mu\nu}$,
the equations of motion for the Lagrangian (\ref{Lc1c2}),
being a special case of Eq. (\ref{MotionEq2}) or
(\ref{EoMundGR}), take the form
\bea
E^\nu_{(2)\mu}&=&R^\nu_{\mu}+2c_1RR^\nu_{\mu}
+2\tilde{c}_2R_{\mu\rho}^{\nu\sigma}R^{\rho}_{\sigma}
-\frac{1}{2}L_2\delta^\nu_{\mu} \nonumber  \\
&=&-\frac{D-4}{D-2}(D-1)(Dc_1+\tilde{c}_2)
\hat{\Lambda}^2\delta^\nu_{\mu}
\nn  \\
&&-\big(\hat{\Lambda}-\hat{\Lambda}_{gr}\big)\delta^\nu_{\mu}
\, .  \label{MotEqL2}
\eea
Substituting Eqs. (\ref{L2forRic}) and (\ref{Tildc2})
into Eq. (\ref{MotEqL2}), we obtain further $E^\nu_{(2)\mu}=0$.
This implies that the general relativistic solutions fulfilling
$R_{\mu\nu}=(D-1)\hat{\Lambda}g_{\mu\nu}$ must be the ones of the
field equations derived from the Lagrangian (\ref{Lc1c2}), which is
the usual Einstein-Hilbert Lagrangian supplemented with the quadratic
curvature terms $R^2$ and $R^{\alpha\beta}R_{\alpha\beta}$.
For instance, the $D$-dimensional Schwarzschild-AdS black holes
in general relativity are solutions of the quadratic
gravity described by the Lagrangian (\ref{Lc1c2}) \cite{NojOd01}.
In addition, when $\hat{\Lambda}=0$, the quadratic-curvature
Lagrangian associated with the solutions satisfying
$R_{\mu\nu}=0$ in any dimension can be expressed as
\begin{equation}
\sqrt{-g}L_3=\sqrt{-g}\left(R+c_1 R^2
+c_2R^{\alpha\beta}R_{\alpha\beta}\right)
\, . \label{L2noL}
\end{equation}
Apparently, since the $D$-dimensional Kerr-AdS black hole
given by Eq. (\ref{MeDKAdS}) fulfills the field equations
$R_{\mu\nu}=-(D-1)\ell^2 g_{\mu\nu}$ derived from the Einstein-Hilbert
Lagrangian $\mathcal{L}_{EH}=\sqrt{-g}[R+(D-1)(D-2)\ell^2]$
\cite{GLuPP1,GLuPP2}, the expression (\ref{MotEqL2}) for field
equations supports that such a black hole solution is also an exact
one for the Lagrangian (\ref{Lc1c2}) with $\hat{\Lambda}=-\ell^2$.
Apart from this, it was shown in \cite{Cl5Dsol} that  a solution
in three-dimensional new massive gravity with constant scalar
curvature can be embedded into five-dimensional quadratic gravity
through dimension lifting.

Subsequently, we move on to demonstrate that the Lagrangian (\ref{Lc1c2})
includes the one describing $D$-dimensional critical gravity
proposed in \cite{KKSCriG}. To do this, we let both the arbitrary
constants $c_1$ and $\hat{\Lambda}$ in the Lagrangian (\ref{Lc1c2})
take the following values
\bea
c_1&=&-\frac{D^2}{4(D-1)(D-2)^3\hat{\Lambda}_{gr}}
=-\frac{D^2}{8(D-2)^2\Lambda}\, , \nonumber \\
\hat{\Lambda}&=&\frac{2(D-2)}{D}\hat{\Lambda}_{gr}
=\frac{4\Lambda}{D(D-1)}
\, , \label{effLinCG}
\eea
respectively, which obey the constraint (\ref{HatLamcons}) to
the constant parameters. Then we arrive at the Lagrangian (\ref{Lc1c2})
for the higher-dimensional critical gravity \cite{KKSCriG}, that is,
\bea
\mathcal{L}_{CG}&=&\sqrt{-g}\left[R-2\Lambda
-\frac{D^2}{8(D-2)^2\Lambda}\left(R^2
-\frac{4D-4}{D}R_{\rho\sigma}R^{\rho\sigma}\right)\right]\nn \\
&=&\sqrt{-g}\left[R-2\Lambda
-\frac{D(D-1)}{8(D-2)(D-3)\Lambda}\left(L_{GB}
-C^{\mu\nu}_{\rho\sigma}C_{\mu\nu}^{\rho\sigma}\right)\right]
\, . \label{LCriGinD}
\eea
The last expression in Eq. (\ref{LCriGinD}) can be acquired directly
by setting $c_1=-D^2/[8(D-2)^2\Lambda]$, $c_2=-4(D-1)c_1/D$ and
$c_3=0$ in the Lagrangian (\ref{QuaCLag2}). It reveals that critical
gravity can be understood as the linear combination of
Einstein-Gauss-Bonnet gravity and Weyl gravity.
According to Eq. (\ref{MotionEq2}), the expression of the equations
of motion for the $D$-dimensional critical gravity is read off as
\bea
E^{(DDCG)}_{\mu\nu}&=&E^{(gr)}_{\mu\nu}
-\frac{D}{16\Lambda(D-2)^2}\Big[4DRR_{\mu\nu}
-16(D-1)R_{\mu\rho\nu\sigma}R^{\rho\sigma}
\nonumber  \\
&&+\Big(4(D-1)R_{\rho\sigma}R^{\rho\sigma}
-D R^2+4\Box R\Big)g_{\mu\nu}  \nonumber  \\
&&-8(D-1)\Box R_{\mu\nu}
+4(D-2)\nabla_\mu\nabla_\nu R \Big]
\, .  \label{MoEqDDCG}
\eea
When $\Lambda=-D(D-1)\ell^2/4$ or $\hat{\Lambda}=-\ell^2$, it can be
verified that the $D$-dimensional Kerr-AdS black hole solution
given by Eq. (\ref{MeDKAdS}) indeed obeys the field equations
$E^{(DDCG)}_{\mu\nu}=0$.

At last, let us make some remarks on the Lagrangian
$\sqrt{-g}L_2$. Such a Lagrangian has three free
parameters denoted by
$(\Lambda,c_1,\hat{\Lambda})$ or $(\Lambda,c_1,\tilde{c}_2)$.
It covers the Lagrangians for Einstein gravity
$(c_1=0,\hat{\Lambda}=\hat{\Lambda}_{gr})$ and
higher-dimensional critical gravity ($c_1$ and
$\hat{\Lambda}$ are given by Eq. (\ref{effLinCG})),
which only possess one
parameter. In fact, within the framework of quadratic
gravities, there exist other single-parameter
Lagranians that admit the solutions satisfying
$R_{\mu\nu}=(D-1)\hat{\Lambda} g_{\mu\nu}$, such as
\bea
\sqrt{-g}L_2|_{\Lambda,\tilde{c}_2=0}&=&\sqrt{-g}
\left(R-\frac{D-2}{D(D-1)(D-4)\hat{\Lambda}} R^2
\right) \, , \nn \\
\sqrt{-g}L_2|_{\Lambda,c_1=0}&=&
\sqrt{-g}\left(R-\frac{D-2}{(D-1)(D-4)\hat{\Lambda}}
R^{\alpha\beta}R_{\alpha\beta}\right)
\, , \label{OnePaLag}
\eea
together with the following Lagrangian
\be
\tilde{\mathcal{L}}^{(DD)}_{CG}=\sqrt{-g}
(L_2-R+2\Lambda)\big|_{\hat{\Lambda}=\hat{\Lambda}_{gr}}
=c_1\sqrt{-g}\big(R^2-D R^{\alpha\beta}R_{\alpha\beta}\big)
\, , \label{DDCriG2}
\ee
which coincides with the Lagrangian (\ref{4DCriG2})
for four-dimensional critical gravity proposed
in \cite{EdNaCG} when $D=4$ and $c_1=-3\alpha$,
arising from that the Gauss-Bonnet term in
four dimensions is non-dynamical.
Due to the fact that the potential for the
Lagrangian (\ref{DDCriG2}), being of the form
\be
\tilde{K}^{\mu\nu}_{DDCG}
=2c_1\Big(R\nabla^{[\mu}\xi^{\nu]}
+DR^{[\mu}_\rho\nabla^{\nu]}\xi^{\rho}
+D\xi^\rho\nabla^{[\mu}R^{\nu]}_\rho\Big)
-c_1(D-4)\xi^{[\mu}\nabla^{\nu]} R
\, , \label{KmmDDCG2}
\ee
vanishes identically for the solutions obeying
$R_{\mu\nu}=(D-1)\hat{\Lambda} g_{\mu\nu}$,
$\tilde{\mathcal{L}}^{(DD)}_{CG}$ may be
interpreted as the higher-dimensional
generalization for the Lagrangian (\ref{4DCriG2}).
Besides, there also exist Lagrangians with double
free parameters, for instance,
\bea
\sqrt{-g}L_2|_{\tilde{c}_2=0}&=&
\sqrt{-g}\left(R-2\Lambda
+\frac{2\Lambda-(D-1)(D-2)\hat{\Lambda}}{D(D-4)(D-1)^2\hat{\Lambda}^2}
R^2\right) \, ,  \nn \\
\sqrt{-g}L_2|_{c_1=0}&=&
\sqrt{-g}\left(R-2\Lambda
+\frac{2\Lambda-(D-1)(D-2)\hat{\Lambda}}{(D-4)(D-1)^2\hat{\Lambda}^2}
R^{\alpha\beta}R_{\alpha\beta}\right)
\, , \label{TwoPaLag}
\eea
as well as the Lagrangian without any coupling parameter,
namely,
$\sqrt{-g}\big(R^2-D R^{\alpha\beta}R_{\alpha\beta}\big)$.
What is more, maybe one expects to acquire a more
general Lagrangian than $\sqrt{-g}L_2$
that admits the solutions fulfilling
$R_{\mu\nu}=(D-1)\hat{\Lambda} g_{\mu\nu}$
by incorporating the square of the Riemann tensor
$R^{\rho\sigma}_{\alpha\beta}
R^{\alpha\beta}_{\rho\sigma}$ within the
Lagrangian. In doing so, one could take
into consideration of the Lagrangian $\sqrt{-g}L$.
For such a Lagrangian, apart from the necessary condition
$R_{\mu\nu}=(D-1)\hat{\Lambda} g_{\mu\nu}$, the
field equations given by Eq. (\ref{EoMundGR})
yields another constraint
for the metric tensor $g_{\mu\nu}$, that is,
\be
\frac{c_3}{D-1}
\big(4R^{\rho\sigma}_{\mu\lambda}R^{\nu\lambda}_{\rho\sigma}
-\delta^\nu_\mu R^{\rho\sigma}_{\alpha\beta}
R^{\alpha\beta}_{\rho\sigma}\big)
=(D-1)(D-4)(Dc_1+c_2)\hat{\Lambda}^2\delta^\nu_\mu
+(D-2)\big(\hat{\Lambda}-\hat{\Lambda}_{gr}\big) \delta^\nu_\mu
\, . \label{RsoconstR}
\ee
Substituting $c_2=\tilde{c}_2$ into the above equation leads
to $c_3=0$ or $4R^{\rho\sigma}_{\mu\lambda}
R^{\nu\lambda}_{\rho\sigma}
=\delta^\nu_\mu R^{\rho\sigma}_{\alpha\beta}
R^{\alpha\beta}_{\rho\sigma}$ for the Riemann tensor.
Since the latter can not
be always guaranteed to hold under the condition
$R_{\mu\nu}=(D-1)\hat{\Lambda} g_{\mu\nu}$, a simple and
feasible setting is $c_3=0$ to render the solutions
of $R_{\mu\nu}=(D-1)\hat{\Lambda} g_{\mu\nu}$
as the ones of $E_{\mu\nu}=0$. This implies
that the Lagrangian $\sqrt{-g}L_2$ with three
parameters can be regarded as the most general one for
quadratic gravities that embrace solutions only required
to obey $R_{\mu\nu}=(D-1)\hat{\Lambda} g_{\mu\nu}$.
For example, although Einstein-Gauss-Bonnet gravity
$(c_3\neq 0)$ is a natural generalization of
general relativity, the higher-dimensional
Kerr-AdS black hole solutions in the context of the latter
can not be embedded into the former because of the
failure of those solutions to satisfy Eq. (\ref{RsoconstR}).

%%%%%%%%%%%%%%%%%%%%%%%%%%%%%%%%%%%%%%%%%%%%%%%%%%%%%%%%
\section{The first law of thermodynamics for
Kerr-AdS black holes within the framework of
quadratic-curvature gravities} \label{appendC}
%%%%%%%%%%%%%%%%%%%%%%%%%%%%%%%%%%%%%%%%%%%%%%%%%%%%

In the present appendix, we investigate the first law of
Kerr-AdS black holes in the theory of Einstein gravity corrected
by the curvature terms $R^2$ and $R^{\alpha\beta}R_{\alpha\beta}$.

Within the framework of the Einstein gravity theory, the mass $M$
and all the angular momenta $J^{(i)}$s for $D$-dimensional
stationary and axially-symmetric Kerr-AdS black holes
in Eq. (\ref{MeDKAdS}) are given
by \cite{PZL,GibPP,BarnichC,DerKather,GullTek,OleKou}
\bea
M&=&\frac{V_{D-2}}{4\pi}\frac{m}{\prod_{j=1}^n\Xi_j}
\left(\sum_{i=1}^{n}\frac{1}{\Xi_i}-\frac{1-\epsilon}{2}\right)
\, , \nonumber \\
J^{(i)}&=&\frac{V_{D-2}}{4\pi}\frac{ma_i}{\Xi_i\prod_{j=1}^n\Xi_j}
\, . \label{MJKAdS}
\eea
The angular velocities $\Omega_i$ corresponding to the angular
momentum $J^{(i)}$, the entropy $S$ and the surface gravity
$\kappa$ are presented by \cite{GLuPP1,GibPP}
\bea
\Omega_i&=&\frac{a_iH(r_H)}{r_H^2+a_i^2} \, , \quad
S=\frac{V_{D-2}}{4}r_H^{\varepsilon-1}\prod_i^n \frac{r_H^2+a_i^2}{\Xi_i}
\, , \nonumber \\
\kappa&=&r_HH(r_H)\sum_i^n\frac{1}{r_H^2+a_i^2}
-\frac{1}{r_H}\left(\frac{H(r_H)}{2}\right)^{\varepsilon}
\, , \label{AnVTS}
\eea
respectively. In Eq. (\ref{AnVTS}), the outer horizon radius $r_H$ is
the largest root of $V-2m=0$. The Hawking temperature is then read off
as $T=\kappa/(2\pi)$. It has been demonstrated in \cite{GibPP,BarnichC}
that all the quantities in Eqs. (\ref{MJKAdS}) and (\ref{AnVTS}) satisfy
the differential form for the first law of thermodynamics, that is,
\begin{equation}
dM=TdS+\sum_i^n \Omega_i dJ^{(i)}
\, , \label{FirsLaw}
\end{equation}
together with the Smarr formula
\begin{equation}
\frac{D-3}{D-2}M=TS+\sum_i^n \Omega_i J^{(i)}
\, . \label{SmaForm}
\end{equation}

On the other hand, in the context of the Lagrangian (\ref{Lc1c2})
with $\hat{\Lambda}=-\ell^2$, the mass $M_{qc}$ and the angular
momenta $J^{(i)}_{qc}$s for the Kerr-AdS black holes corrected
by quadratic-curvature terms are presented by Eq. (\ref{MassAninDdi}).
By virtue of the following relation
\begin{equation}
\frac{\partial L_2}{\partial R^{\rho\sigma}_{\mu\nu}}
=-\frac{2(D-2)\hat{\Lambda}_{gr}+D\ell^2}{(D-4)\ell^2}
\frac{\partial R}{\partial R^{\rho\sigma}_{\mu\nu}}
\, ,
\end{equation}
we obtain the entropy $S_{qc}$ via the Wald's entropy formula
for black holes \cite{IyW,WalEnt,JaKMy}, being of the form
\begin{equation}
S_{qc}
=-\frac{2(D-2)\hat{\Lambda}_{gr}+D\ell^2}{(D-4)\ell^2}S
\, .\label{entrSqc}
\end{equation}
The angular velocities, the entropy and the surface gravity are still
presented by Eq. (\ref{AnVTS}). Consequently, by letting
Eqs. (\ref{FirsLaw}) and (\ref{SmaForm}) multiplied by the factor
$-[2(D-2)\hat{\Lambda}_{gr}+D\ell^2]/[(D-4)\ell^2]$, respectively,
we obtain the first law of
thermodynamics for the Kerr-AdS black holes in the gravity theory
with quadratic-curvature terms, namely,
\begin{equation}
dM_{qc}=TdS_{qc}+\sum_i^n \Omega_i dJ^{(i)}_{qc}
\, , \label{FirLinQC}
\end{equation}
as well as the following Smarr formula
\begin{equation}
\frac{D-3}{D-2}M_{qc}=TS_{qc}+\sum_i^n \Omega_i J^{(i)}_{qc}
\, . \label{SmaFoQC}
\end{equation}

%%%%%%%%%%%%%%%%%%%%%%%%%%%%%%%%%%%%%%%%%%%%%%%%%%%%%%%%
\section{Comparison between various potentials}
\label{appendD}
%%%%%%%%%%%%%%%%%%%%%%%%%%%%%%%%%%%%%%%%%%%%%%%%%%%%

In Sec. \ref{two}, we put forward the potential
$K^{\mu\nu}$ in Eq. (\ref{Kmunu0}) or (\ref{Kmunu})
for the computation of the
conserved charges of asymptotically AdS spacetimes
in the context of quadratic gravities described by
the Lagrangian $\sqrt{-g}L$ in Eq. (\ref{QuaCLag}).
To illustrate its significance, we also gave some special
cases of $K^{\mu\nu}$, including
$K^{\mu\nu}_{Weyl}$, $K^{\mu\nu}_{EGB}$,
$K^{\mu\nu}_{gr}$ and $K^{\mu\nu}_{Ric}$,
corresponding to the Lagrangians
$\sqrt{-g}L_W$ in Eq. (\ref{WeylLag}),
$\sqrt{-g}L_{EGB}$ in Eq. (\ref{GBLag}),
$\sqrt{-g}L_{EH}=\sqrt{-g}\big[R-(D-1)(D-2)\hat{\Lambda}\big]$
and $\sqrt{-g}L$ constrained by the solutions satisfying
$R_{\mu\nu}=(D-1)\hat{\Lambda}g_{\mu\nu}$, respectively.
We compared those potentials with some existing ones
defined through various methods developed in the
literature, such as the (off-shell generalized)
ADT method, the field-theoretic approach and the
topological regularization method. Some equivalences
to the potentials proposed in this work were presented.
For convenience to see all the results, we summarize
them in Table \ref{PotComp}. For example, in the third
row, the linear perturbation of the two-form
$K^{\mu\nu}_{Weyl}$ on
the AdS spaces, $\delta K^{\mu\nu}_{Weyl}$, is consistent
with the ADT potential given by Eq. (23) in \cite{JJPEPJC}.
In the seventh row, the potential $K^{\mu\nu}($even $D)$
is equivalent to the one given by  Eq. (23)
in \cite{WanPen} and Eq. (14) in \cite{GMORB},
obtained by means of the topological regularization method.

\begin{table}[H]
\caption{Results from comparison between various potentials}
    \vspace{15pt}
    \centering
    \begin{tabular}{p{2.10cm}p{2.20cm}p{5.55cm}}
        \hline
        \hline
     Lagrangian   &Potential  & Equivalences of the potential   \\
        \hline
        $\sqrt{-g}L$   &$\delta K^{\mu\nu}$
        &Eq. (\ref{QbarIW}) via the ADT method  \\
        $\sqrt{-g}L_W$   &$\delta K^{\mu\nu}_{Weyl}$
        &Eq. (23) in \cite{JJPEPJC}   \\
        $\sqrt{-g}L_{EGB}$   &$\delta K^{\mu\nu}_{EGB}$
        &Eq. (3.11) in \cite{PeEGB}, Eq. (41) in \cite{JJPEPJC}   \\
        $\sqrt{-g}L_{EH}$   &$K^{\mu\nu}_{gr}$
        &Eq. (2.15) in \cite{PZL}   \\
        $\sqrt{-g}L$   &$(\delta) K_{Ric}^{\mu\nu}$
        &Eq. (2.3) in \cite{AmGor}, Eq. (6) in \cite{SeST}   \\
        $\sqrt{-g}L$   &$K^{\mu\nu}($even $D)$
        &Eq. (23) in \cite{WanPen}, Eq. (14) in \cite{GMORB}   \\

      \hline
    \end{tabular}
    \label{PotComp}
\end{table}

\end{document}